# Magnetic-field-induced charge-stripe order in the high temperature superconductor $YBa_2Cu_3O_y$

Tao Wu[1], Hadrien Mayaffre[1], Steffen Krämer[1], Mladen Horvatić[1], Claude Berthier[1], W.N. Hardy[2,3], Ruixing Liang[2,3], D.A. Bonn[2,3], and Marc-Henri Julien[1]

*1 Laboratoire National des Champs Magnétiques Intenses, UPR 3228, CNRS-UJF-UPS-INSA, 38042 Grenoble, France*

*2 Department of Physics and Astronomy, University of British Columbia, Vancouver, BC V6T 1Z1, Canada*

*3 Canadian Institute for Advanced Research, Toronto, Ontario M5G 1Z8, Canada*

**Electronic charges introduced in copper-oxide ($CuO_2$) planes generate high-transition temperature ($T_c$) superconductivity but, under special circumstances, they can also order into filaments called stripes (1). Whether an underlying tendency of charges to order is present in all cuprates and whether this has any relationship with superconductivity are, however, two highly controversial issues (2,3). In order to uncover underlying electronic orders, magnetic fields strong enough to destabilise superconductivity can be used. Such experiments, including quantum oscillations (4-6) in $YBa_2Cu_3O_y$ (a notoriously clean cuprate where charge order is not observed) have suggested that superconductivity competes with spin, rather than charge, order (7-9). Here, using nuclear magnetic resonance, we demonstrate that high magnetic fields actually induce charge order, without spin order, in the $CuO_2$ planes of $YBa_2Cu_3O_y$. The observed static, unidirectional, modulation of the charge density breaks translational symmetry, thus explaining quantum oscillation results, and we argue that it is most likely the same 4*a*-** **periodic modulation as in stripe-ordered cuprates (1). The discovery that it develops only when superconductivity fades away and near the same 1/8$^{th}$ hole doping as in $La_{2-x}Ba_xCuO_4$ (1) suggests that charge order, although visibly pinned by CuO chains in $YBa_2Cu_3O_y$, is an intrinsic propensity of the superconducting planes of high $T_c$ cuprates.**

The ortho II structure of $YBa_2Cu_3O_{6.54}$ ($p$=0.108) leads to two distinct planar Cu NMR sites: Cu2F are those Cu located below oxygen-filled chains and Cu2E those below oxygen-empty chains (10). The main discovery of our work is that, on cooling in a field $H_0$ of 28.5 T along the ***c*** axis (i.e. in the conditions for which quantum oscillations are resolved; See supplementary materials), the Cu2F lines undergo a profound change while the Cu2E lines do not (Fig. 1). To first order, this change can be described as a splitting of Cu2F into two sites having both different hyperfine shifts $K=\langle h_z\rangle/H_0$ ($\langle h_z\rangle$ is the hyperfine field due to electronic spins) and quadrupole frequencies $\nu_Q$ (Related to the electric field gradient). Additional effects might be present (Fig. 1), but they are minor compared to the observed splitting. Field and temperature dependent orbital occupancy (*e.g.* $d_{x^2-y^2}$ *vs.* $d_{z^2-r^2}$) changes without on-site electronic density change are implausible and any out-of-plane charge-density or lattice change would affect Cu2E sites as well. Thus, the change of $\nu_Q$ can only arise from a differentiation in the charge density among Cu2F sites (or at the oxygen sites bridging them). A change in the asymmetry parameter and/or in the direction of the principal axis of the electric field gradient could also be associated with this charge differentiation, but these are relatively small effects.

Remarkably, the charge differentiation occurs below $T_{charge}$=50±10 K for $p$=0.108 (Figures 1, S9, S10) and 67±5 K for $p$=0.12 (Figures S7, S8). Within error bars, $T_{charge}$ coincides for each of the samples with $T_0$, the temperature at which the Hall constant $R_H$ becomes negative, an indication of the Fermi surface reconstruction (12-14). Thus, whatever the precise profile of the static charge modulation is, the reconstruction must be related to the translational symmetry breaking by the charge ordered state.

The absence of any splitting or broadening of Cu2E lines implies a one-dimensional character of the modulation within the planes and imposes strong constraints on the charge pattern. Actually, only two types of modulations are compatible with a Cu2F splitting (Fig. 2). The first is a commensurate short-range (2***a*** or 4***a*** period) modulation running along the (chain) ***b***-axis. However, this hypothesis is highly unlikely: To the best of our knowledge, no such modulation has ever been observed in the $CuO_2$ planes of any cuprate, thus it would have to be triggered by a charge modulation pre-existing in the filled chains. A charge-density-wave is unlikely as the finite-size chains are at best poorly conducting in the temperature and doping range discussed here (12,15,16). Any inhomogeneous charge distribution such as Friedel oscillations around chain defects would broaden rather than split the lines. Furthermore, charge order occurs only for high fields perpendicular to the planes as the NMR lines neither split at 15 T nor in a field of 28.5 T parallel to the $CuO_2$ planes (either along ***a*** or ***b***), two situations in which superconductivity remains robust (Fig. 1). This clear competition between charge order and superconductivity is also a strong indication that the charge ordering instability arises from the planes.

The only other pattern compatible with NMR data is an alternation of more and less charged Cu2F rows defining a modulation with a period of four lattice spacings along the ***a*** axis (Fig. 2). Strikingly, this corresponds to the (site-centered) charge stripes found in $La_{2-x}Ba_xCuO_4$ at doping levels near $p$=$x$=0.125 (1). Being a proven electronic instability of the planes, which is detrimental to superconductivity (2), stripe order not only provides a simple explanation of the NMR splitting, but it also rationalises the striking effect of the field. Stripe order is also fully consistent with the remarkable similarity of transport data in $YBa_2Cu_3O_y$ and in stripe-ordered cuprates (particularly the dome-shaped dependence of $T_0$ around $p$=0.12) (12-14). However, stripes must be parallel from plane-to-plane in $YBa_2Cu_3O_y$, while they are perpendicular in *e.g.* $La_{2-x}Ba_xCuO_4$. We hypothesise that this explains why the charge transport along the ***c***-axis in $YBa_2Cu_3O_y$ becomes coherent in high fields below $T_0$ (16). If so, stripe fluctuations must be involved in the incoherence along ***c*** above $T_0$.

Knowing the doping dependence of $\nu_Q$ (17), the difference $\Delta\nu_Q$=320±50 kHz for $p$=0.108 implies a charge density variation as small as $\Delta p$=0.03±0.01 hole between Cu2Fa and Cu2Fb. A canonical stripe description ($\Delta p$=0.5 hole) is thus inadequate at the NMR timescale of ~$10^{-5}$ s where most (below $T_0$) or all (above $T_0$) of the charge differentiation is averaged out by fluctuations faster than $10^5$ $s^{-1}$. This should not be a surprise: The metallic nature of the compound at all fields is incompatible with full charge order, even if this order is restricted to the direction perpendicular to the stripes (18). Actually, there is compelling evidence of stripe fluctuations down to very low temperatures in stripe-ordered cuprates (19) and indirect evidence (explaining the rotational symmetry breaking) over a broad temperature range in $YBa_2Cu_3O_y$ (15,20-23). Therefore, instead of being a defining property of the ordered state, the small amplitude of the charge differentiation is more likely a consequence of stripe order (the smectic phase of an electronic liquid crystal (18)) remaining partly fluctuating (*i.e.* nematic).

In stripe cuprates, charge order at $T$=$T_{charge}$ is always accompanied by spin order at $T_{spin}$<$T_{charge}$. Slowing down of the spin fluctuations strongly enhances the spin-lattice ($1/T_1$) and spin-spin ($1/T_2$) relaxation rates between $T_{charge}$ and $T_{spin}$ for $^{139}$La nuclei. For the more strongly hyperfine-coupled $^{63}$Cu, the relaxation rates become so large that the Cu signal is gradually 'wiped out' on cooling below $T_{charge}$ (19,24,25). In contrast, the $^{63}$Cu(2) signal here in $YBa_2Cu_3O_y$ does not experience any intensity loss and $1/T_1$ does not show any peak or enhancement as a function of temperature (Fig. 3). Moreover, the anisotropy of the linewidth (Supplementary section II) indicates that the spins, although staggered, align mostly along the field (i.e. ***c***-axis) direction and the typical width of the central lines at base temperature sets an upper magnitude for the static spin polarisation as small as $g<S_z> \leq 2\ 10^{-3}\ \mu_B$ for both samples in fields of ~30 T. These consistent observations rule out the presence of magnetic order, as previously suggested (6).

In stripe-ordered cuprates, the strong increase of $1/T_2$ on cooling below $T_{charge}$ is accompanied by a crossover of the time-decay of the spin-echo from the high temperature Gaussian form $\exp(-1/2(t/T_{2G})^2)$ to an exponential form $\exp(-t/T_{2E})$ (19,24). A similar crossover occurs here, albeit in a less extreme manner because of the absence of magnetic order: $1/T_2$ sharply increases below $T_{charge}$ and the decay actually becomes a combination of exponential and Gaussian decays (Figure 3). In supplementary section VII, we provide evidence that the typical values of the $1/T_{2E}$ below $T_{charge}$ imply that antiferromagnetic (or 'spin-density-wave') fluctuations are slow enough to appear frozen on the timescale of a cyclotron orbit $1/\omega_c \approx 10^{-12}$ s. In principle, such slow fluctuations could reconstruct the Fermi surface, provided spins are correlated over large enough distances (26,27) (See also 9). It is unclear whether this condition is fulfilled here. The fluctuations could also appear frozen on the timescale of an elastic neutron scattering experiment, as in $YBa_2Cu_3O_{6.45}$ ($p \approx 0.08$) (8,20). However, there is a fundamental difference between $p$=0.08 and the $p$=0.108-0.12 samples here, which is not a question of experimental timescale: Our (unpublished) NMR data in $YBa_2Cu_3O_{6.45}$ are completely different from those reported here, with unequivocal evidence of spin order. Even if we cannot exclude that a freezing at the NMR timescale occurs at much lower temperatures and/or higher field, spin order appears to be absent over a range of temperatures and fields where charge order and quantum oscillations are observed, therefore indicating that it cannot be an essential ingredient of these two phenomena. Actually, the phase diagram of underdoped $YBa_2Cu_3O_y$ (Fig. 4) even suggests that AFM order (28) and Fermi surface reconstruction are mutually exclusive phenomena. It is tempting to associate the absence of spin order to the remaining stripe fluctuations discussed above.

The implications of our results go beyond the microscopic explanation of quantum oscillation experiments. While it is the chain structure which manifestly pins stripe order in $YBa_2Cu_3O_y$, the chains should not be taken as responsible for the whole stripe phenomenon here. First, charge order onsets near $T_0$ where the Hall effect changes its sign, and this effect has been shown to be the same planar '1/8 anomaly' as in stripe-ordered cuprates (13,14). Second, $T_0$ (Ref.13 and Fig. 4) is a continuous function of hole doping, irrespective of the oxygen ordering sequence. In particular, the highest $T_{charge}$ is found for the ortho VIII structure ($p$=0.12) which is more complex and shorter range than ortho II ($p$~0.1). Therefore, neither the chains nor the strong disorder typical of La-based stripe cuprates can be the sole origin of charge stripes in ultra clean $YBa_2Cu_3O_y$. Stripe correlations originating from the superconducting planes are actually consistent with the field-tuned competition between charge order and superconductivity revealed here. Our observation of unidirectional charge order in $YBa_2Cu_3O_y$ ($p$=0.11-0.12) thus strengthens the idea that there is an intrinsic and most likely ubiquitous (29,30) propensity of the charges to order in high $T_c$ cuprates, that is most apparent around $p$=1/8.

Acknowledgements: We are grateful to C. Proust for discussions and invaluable help at several stages of this project, as well as to L. Taillefer and S. Kivelson for important discussions. We also thank P. Bourges, P. Carretta, S. Chakraverty, W. Chen, P. Hirschfeld, D. LeBoeuf, A. Millis, M. Norman, B. Ramshaw, S. Sachdev, S. Sanna, M. Takigawa and B. Vignolle for helpful communications.

This work was supported by the Université Joseph Fourier – Grenoble I (pôle SMIng) and Euromagnet II.

Authors' Contributions: W.N.H., R.L. and D.A.B. prepared the samples. T.W., H.M, S.K. and M.-H.J. performed the experiments. S.K. and M.H. developed and operate the high field NMR facility. H.M. created software for spectrometers and data analysis. T.W. and M.-H.J. analyzed the data. C.B. provided conceptual advice and contributed to the planning of the project. M.H.J. wrote the paper and supervised the project. All authors discussed the results and commented on the manuscript.

The authors declare no competing financial interests. Correspondence and requests for materials should be addressed to M.-H.J. (marc-henri.julien@lncmi.cnrs.fr).

**Figure 1 | High field NMR spectra of $YBa_2Cu_3O_{6.54}$ (ortho-II, p=0.108).**

**a**, $^{63}$Cu(2) NMR lines for $H_0||\mathbf{c}$ (15 T) do not show any temperature-induced splitting. A small overall shift has been subtracted for clarity. The slight change of shape of the central line at 1.5 K is due to Cu2F broadening and $T_2$ shortening but the Cu2E/Cu2F positions are unchanged with respect to higher temperatures and lower fields. A small background on the central line at 58 K has been subtracted (See supplementary section IV). The $\nu_Q$ difference between Cu2E and Cu2F is due to the presence of empty and filled chains creating different local environments, minute differences in the orbital occupations and, possibly, slightly different charge densities at these sites (This later possibility is yet unsettled but this is not directly relevant to the present work as it is field independent up to at least 300 K). **b**, Cu2F lines for $H_0||ab$ (28.5 T) show no splitting or broadening at 1.5 K. Cu2E satellites are not shown. **c**, $^{63}$Cu lines for $H_0||c$ (28.5 T). **d,e**, decomposition of the spectrum into the different sites. The charge density modulation below $T_{charge}$≈50 K, causes Cu2F splitting to split into Cu2Fa and Cu2Fb (higher density site for its larger $\nu_Q$ value (17)). A similar NMR splitting might have been observed in stripe ordered nickelates (11). The parameters are given in supplementary table S1. The central line positions are determined from the positions of the satellites, and they are independently confirmed in Supplementary figures S4 and S5. Stars indicate signals from $^{63}$Cu in oxygen-empty chains. A subtle bump on the low frequency side of both Cu2F and Cu2E high frequency satellites is not understood, but is a minor effect compared to the splitting of the Cu2F site and it could be either intrinsic (weak additional magnetic or charge modulation) or extrinsic (defects in the ordered pattern). **f**, quadrupolar contribution to the splitting of the high frequency Cu2F satellite (See figure S9 for details). Error bars include the statistical error from the fit and the uncertainty on the presence of a splitting at high temperatures.

**Figure 2 | Charge density modulations compatible with NMR spectra.**

**a**, 2***b***-periodic charge modulation along the chain (***b***) axis: $p(x,y)=p_0+p_1\sin(\pi y/2)\cos(\pi x)$ with $p_0=0.108$ the mean hole content and $2p_1=0.03$ the amplitude of the modulation. Cu2F (Cu2E) sites lie at odd (even) positions on the ***a***-axis. A similar 4***b***-period modulation of the form $p(x,y)=p_0+p_1\sqrt{2}\,\sin(\pi y/2)\cos(\pi x/2+\pi/4)$ is also possible. These two modulations, though consistent with the NMR spectra, are unable to explain the field dependence of the modulation as well as the correlation between our NMR data, transport measurements and '1/8 anomalies'. **b**, Stripe order with period 4***a*** perpendicular to the chain (***b***) axis: $p(x,y) = p_0+p_1\sin(\pi y/2)$. In order to minimise Coulomb repulsion, the doped holes tend to align below the neutral (filled) chains rather than below the $Cu^+$ (empty) chains, explaining why the higher charge density sites are the Cu2F rather than the 2E sites. Note that, in either case, an additional transverse modulation $p(y)=p_2\cos(\pi y)$ resulting from the ortho II potential and present at any temperature and field cannot be excluded. In the more complex ortho VIII structure of the $p=0.12$ sample, it is also possible to obtain the same striped charge pattern with the high density sites only below filled chains.

**Figure 3 | Slow spin fluctuations instead of spin order.**

**a-b**, temperature dependence of the planar $^{63}$Cu spin-lattice relaxation rate $1/T_1$. The absence of any peak/enhancement on cooling rules out the occurrence of a magnetic transition. **c-d**, increase of the $^{63}$Cu spin-spin relaxation rate $1/T_2$ on cooling below ~$T_{charge}$, obtained from a fit of the spin-echo decay to a stretched form $s(t) \propto \exp(-(t/T_2)^{\alpha})$. **e-f**, stretching exponent $\alpha$. The deviation from $\alpha$=2 on cooling arises mostly from an intrinsic combination of Gaussian and exponential decays, combined with some spatial distribution of $T_2$ values (Supplementary information). The grey areas define the crossover temperature $T_{slow}$ below which slow spin fluctuations cause $1/T_2$ to increase and to become field-dependent; note that the change of shape of the spin-echo decay occurs at slightly higher (~ +15 K) temperature than $T_{slow}$. $T_{slow}$ is slightly lower than $T_{charge}$, consistent with the slow fluctuations being a consequence of charge-stripe order. The increase of $\alpha$ at the lowest temperatures probably signifies that the condition $\gamma \langle h_z^2 \rangle^{1/2} \tau_c \ll 1$ is no longer fulfilled, so that the associated decay is no longer a pure exponential. We note that the upturn of $1/T_2$ is already present at 15 T, while no line splitting is detected at this field. The field thus affects quantitatively, but not qualitatively, the spin fluctuations. **g**, NMR signal intensity (corrected for a temperature factor $1/T$ and for the $T_2$ decay) *vs*. temperature. The absence of any intensity loss at low temperatures also rules out the presence of magnetic order with any significant moment. Error bars represent the added uncertainties in signal analysis, experimental conditions and $T_2$ measurements.

**Figure 4 | Phase diagram of underdoped $YBa_2Cu_3O_y$**

The charge ordering temperature $T_{charge}$ (defined as the onset of the Cu2F line-splitting) coincides with $T_0$, the temperature at which the Hall constant $R_H$ changes its sign. $T_0$ is considered as the onset of the Fermi surface reconstruction (12-14). The continuous line represents the superconducting transition temperature $T_c$. The dashed line indicates the speculative nature of the extrapolation of the field-induced charge order. The magnetic transition temperatures ($T_{spin}$) are from μSR data (28). Remarkably, $T_0$ and $T_{spin}$ vanish close to the same critical concentration $p$=0.08. A scenario of field-induced spin order has been anticipated for $p$>0.08 (8) in analogy with $La_{1.855}Sr_{0.145}CuO_4$ for which the non magnetic ground state switches to AFM order in fields greater than a few Tesla (7 and references therein). Our work, however, shows that spin order does not occur up to ~30 Tesla. On the other hand, the field-induced charge order reported here raises the question of whether a similar field-dependent charge order actually underlies the field dependence of the spin order in $La_{2-x}Sr_xCuO_4$ and $YBa_2Cu_3O_{6.45}$. Error bars represent the uncertainty in defining the onset of the NMR line splitting (Figures 1F and S8-S10).

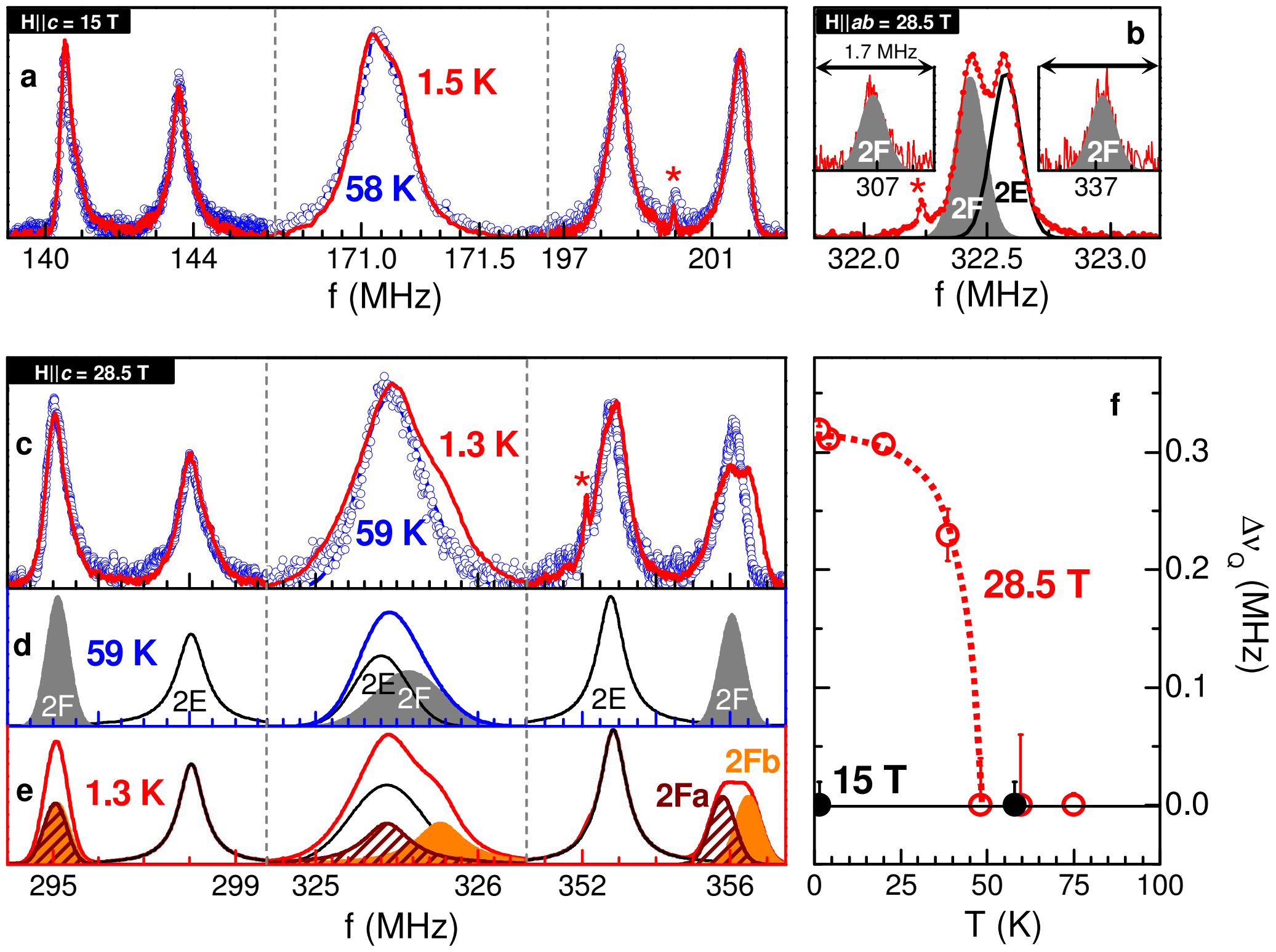

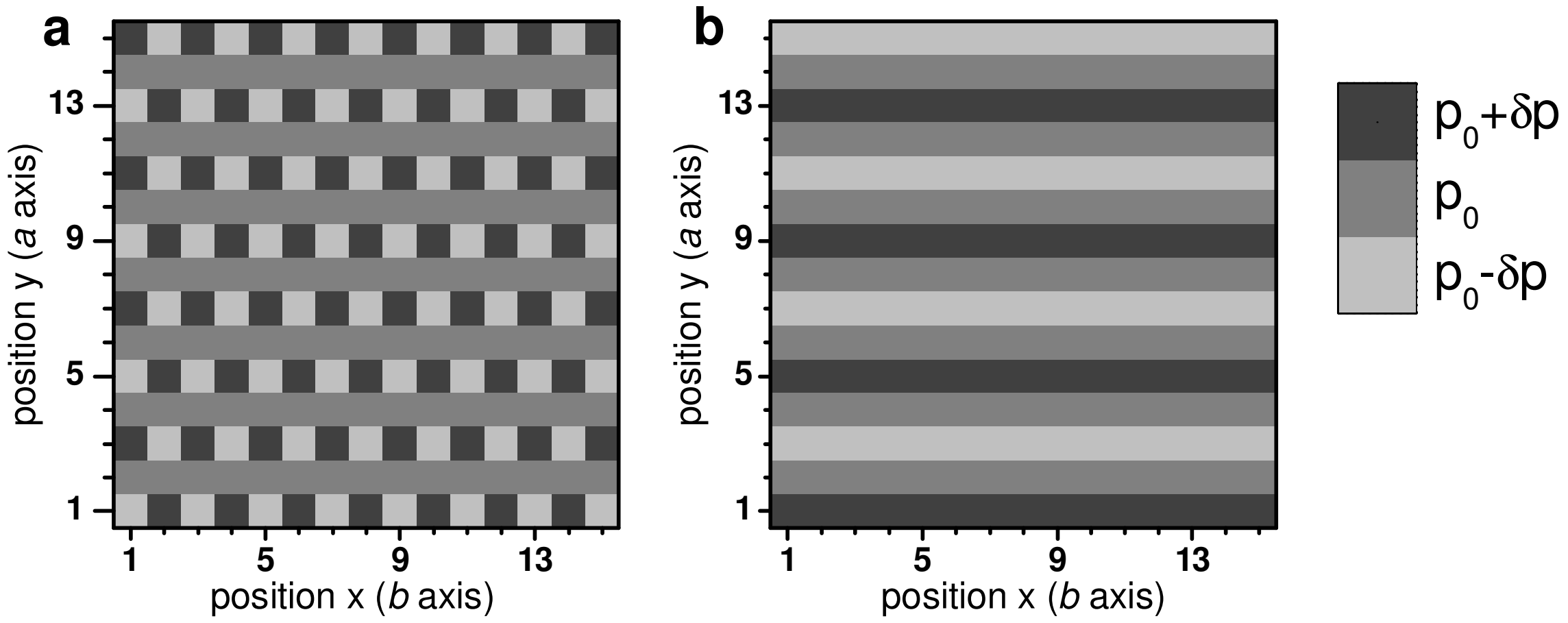

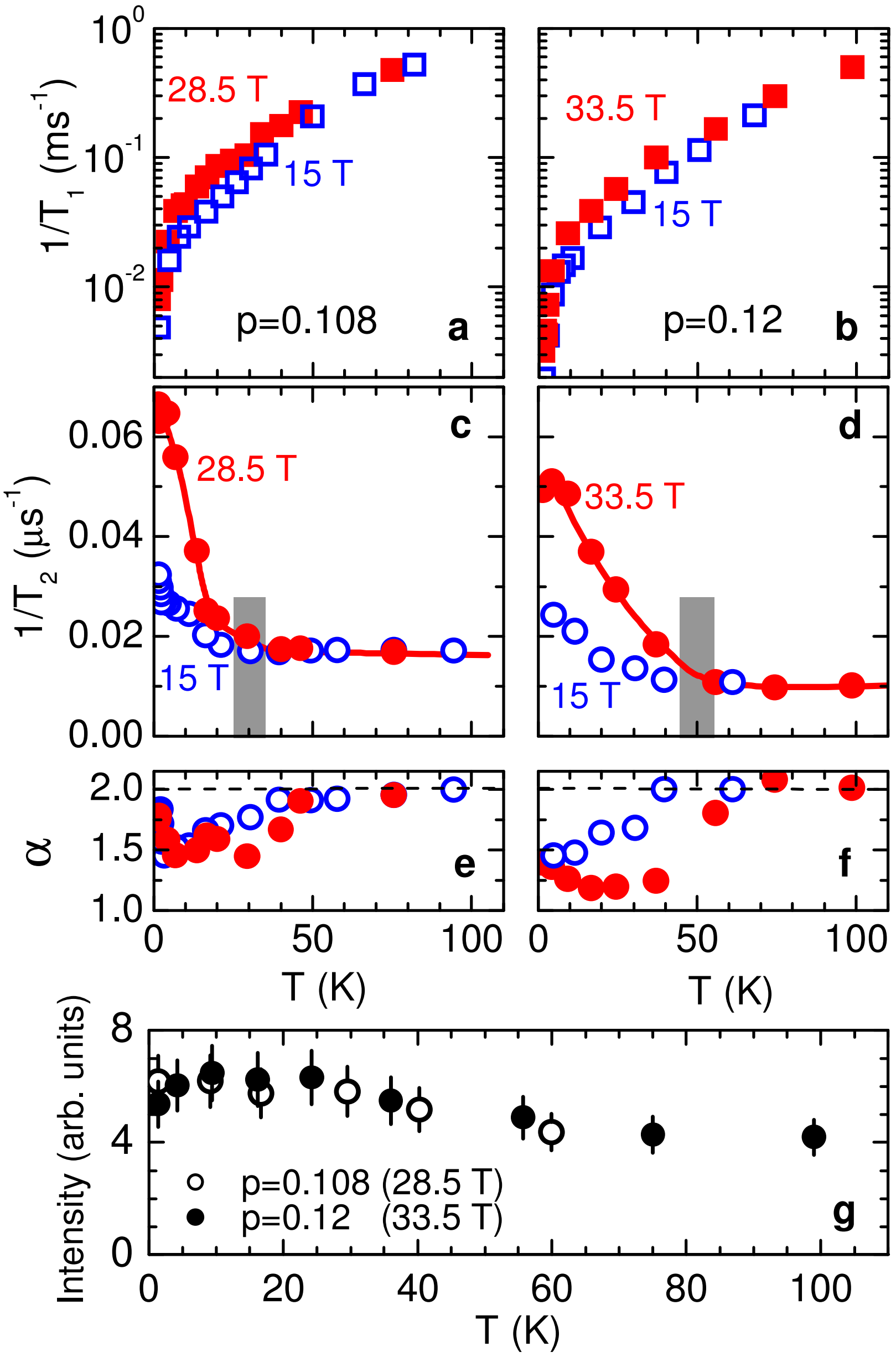

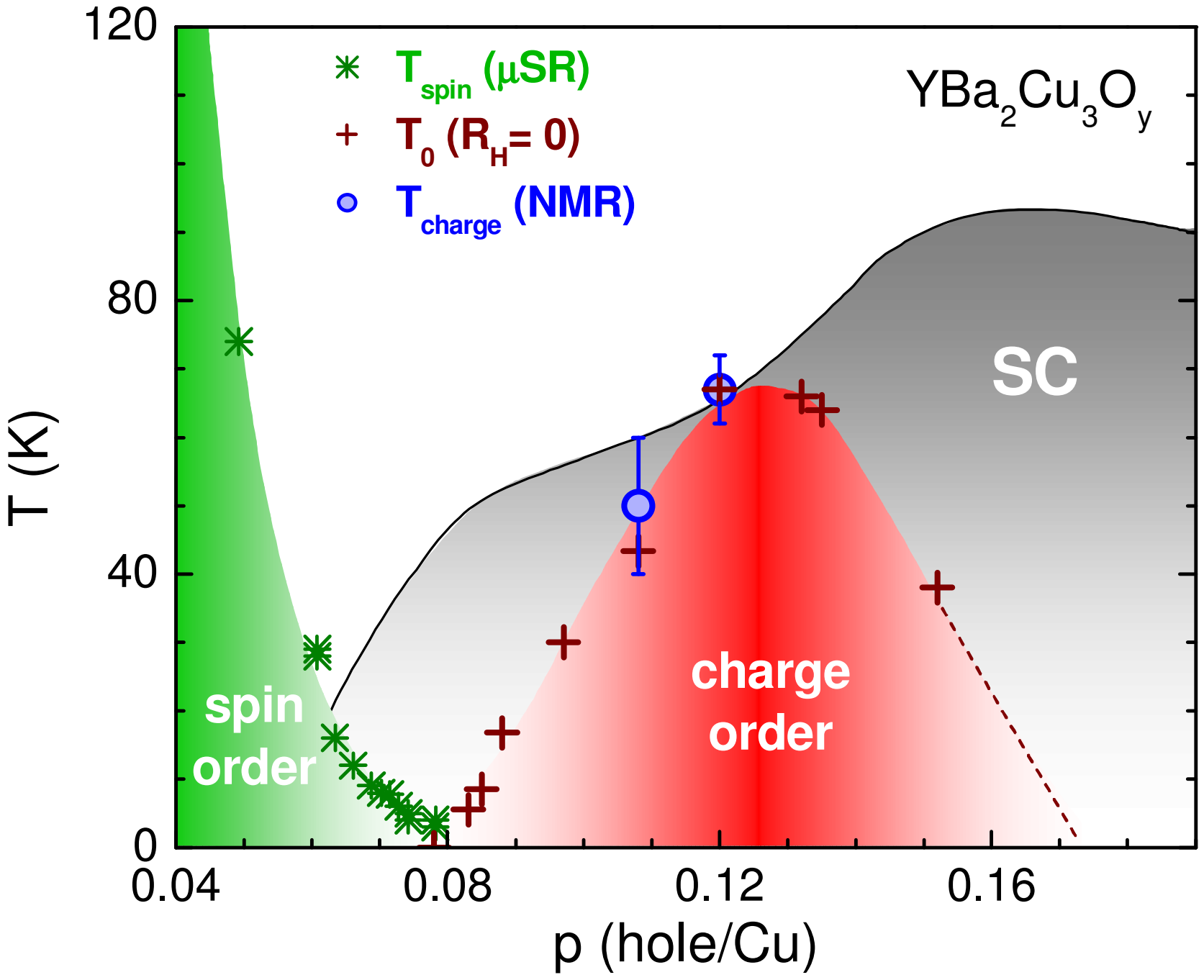

# Supplementary information

## I. Experimental details

**Samples**. Fully detwinned crystals of $YBa_2Cu_3O_y$ were grown in non-reactive $BaZrO_3$ crucibles from high-purity starting materials (see ref. 31 and references therein). For y=6.54 (or 6.67), the dopant oxygen atoms were made to order into an ortho-II (or ortho-VIII) superstructure, yielding a superconducting transition temperature $T_c$=61.3 K (or 66.0 K). The samples are uncut, unpolished platelets of size 0.9 x 0.8 x 0.1 $mm^3$ (or 1.3 x 1.2 x 0.3 $mm^3$). Quantum oscillations were measured in the $YBa_2Cu_3O_{6.54}$ sample studied here (32).

**Conditions**. The conditions to observe quantum oscillations in $YBa_2Cu_3O_y$ are: oxygen content y~6.5-6.7 (*i.e.* 0.085-0.14 hole per Cu), magnetic fields $H_0$ above ~24 T applied along the ***c***-axis (14) and temperatures *T* below ~18 K (33). Still, the Fermi-surface reconstruction itself is considered to take place at even lower fields and higher temperatures (12-14), but the oscillations are too strongly damped to be observed. Static magnetic fields above 15 T were generated by the M9 resistive magnet of the LNCMI Grenoble.

**NMR.** Standard spin-echo techniques were used with a home-built heterodyne spectrometer. Spectra were obtained at fixed magnetic fields by combining Fourier transforms of the spin-echo signal recorded for regularly-spaced frequency values (34). The $^{27}$Al NMR signal from a 0.8 μm-thick aluminium foil (99.1 % purity) was used to calibrate both the external field and the absolute NMR intensities. The NMR intensity data were corrected from the temperature-dependent spin-echo decay. $T_1$ values were obtained from standard fits to the recovery of the nuclear magnetisation of a nuclear spin I=3/2 after saturation or inversion. A stretching exponent was added to account for the distribution of $T_1$ values, mainly showing up at low temperatures. Because $^{63}$Cu and $^{65}$Cu have a nuclear spin I=3/2 each site shows six resonance lines, three per isotope: a central line (-1/2,1/2 transition), as well as low frequency (-3/2,-1/2 transitions) and high frequency (1/2,3/2 transitions) satellites, split by the quadrupole interaction.

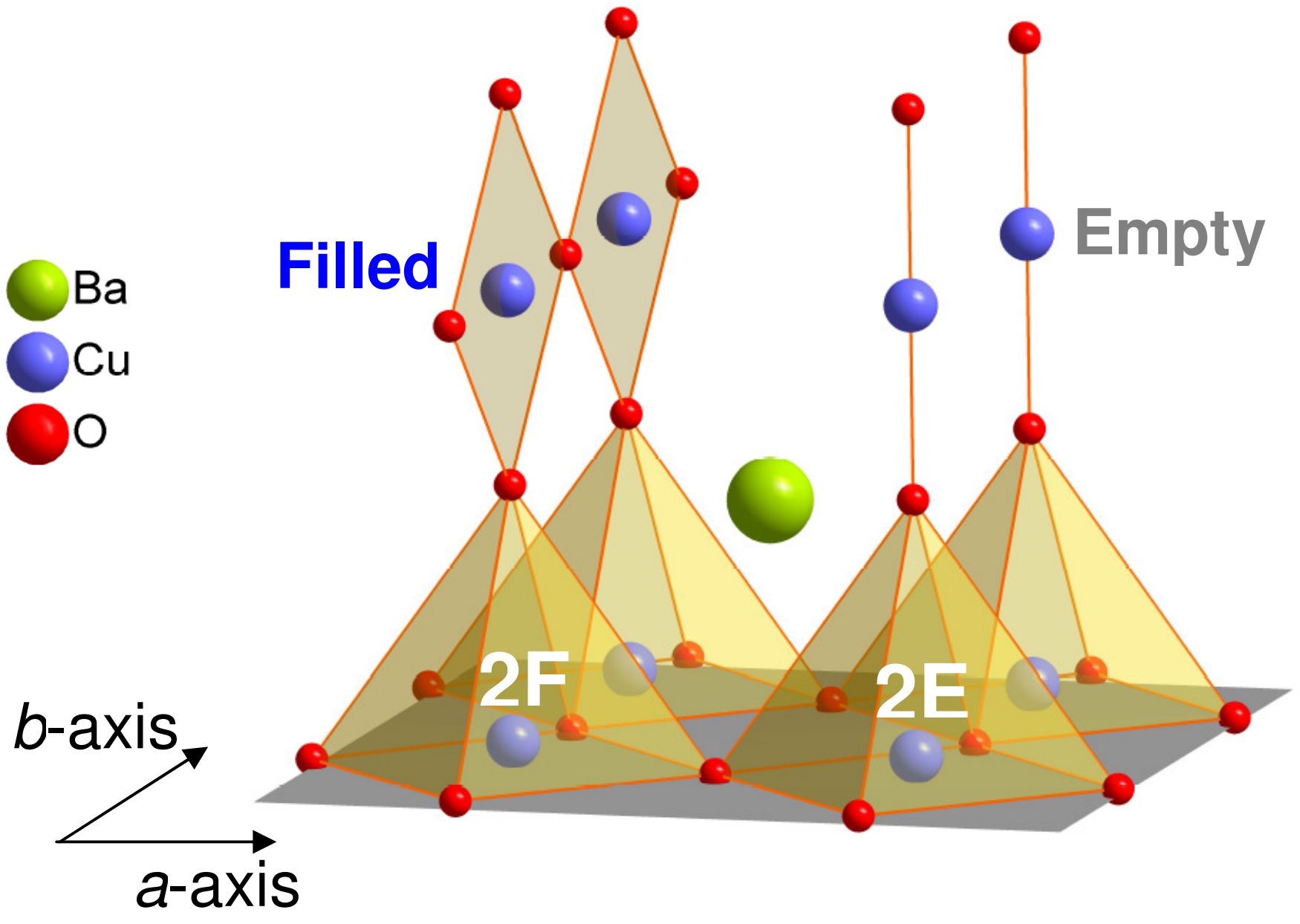

**Figure S1 | Structure of ortho-II YBCO**. The alternating sequence of oxygen-filled and oxygen-empty chains is responsible for the existence of two different planar (Cu2E and Cu2F) copper sites. These have different hyperfine fields and quadrupole frequencies at least up to room temperature.

# II. Hyperfine fields and staggered magnetisation at Cu2 sites

In high-$T_c$ cuprates, the magnetic hyperfine field $\boldsymbol{h}$ at each planar Cu(2) site $(i,j)$ is related to the on-site electronic spin polarisation $<S_z(i,j)>$ but also to the polarisation of the four first neighbours through :

$$<\boldsymbol{h}_{\alpha\alpha}> = g_{\alpha\alpha} (\mathbf{A}_{\alpha\alpha} <S_z(i,j)> + \mathbf{B} <S_z(i\pm1,j)> + \mathbf{B} <S_z(i,j\pm1)>) , \qquad (1)$$

where **A** and **B** are the on-site and transferred hyperfine coupling constants, respectively, and $\alpha = a,b$ or $c$ is the crystallographic direction along which the magnetic field is aligned. For a homogeneous paramagnetic system $<S_z(i,j)>=<S_z(i\pm1,j)>= <S_z(i,j\pm1)>$, so equation (1) reduces to

$$<\boldsymbol{h}_{\alpha\alpha}>=g_{\alpha\alpha} (\mathbf{A}_{\alpha\alpha} + 4\mathbf{B}) <S_z> , \qquad (2)$$

the standard expression used in high $T_c$ cuprates. For H||$c$-axis in $YBa_2Cu_3O_y$, $A_c+4B\approx0$ (actually very slightly negative), so there is in principle no information on the spin polarisation in this field orientation. However, large T-dependent broadening of the Cu2 NMR line shows that Eqn. (2) is not satisfied for many sites because the spin polarisation is spatially inhomogeneous (10,35). Furthermore, the much larger broadening for H||$c$ than for H||$ab$ (Fig. S2) can be understood only if the local polarisation is essentially staggered and aligned along the field direction (See Refs. 36,37).

Since the spatial profile of the staggered polarisation is not accurately known both in the absence and in the presence of stripe order, we assume for simplicity that the modulus $|<Sz>|$ is spatially uniform, while values are staggered i.e. $<S_z(i,j)> = -<S_z(i\pm1,j)> = -<S_z(i,j\pm1)>$. In these conditions and neglecting the non-staggered component of the polarisation, we rewrite $<\boldsymbol{h}_{\alpha\alpha}>=g_{\alpha\alpha} (\mathbf{A}_{\alpha\alpha} - 4\mathbf{B}) <S_z(i,j)>$. For numerical estimates, we use $A_c$=-155 kG/$\mu_B$ and B=38 kG/$\mu_B$ (38).

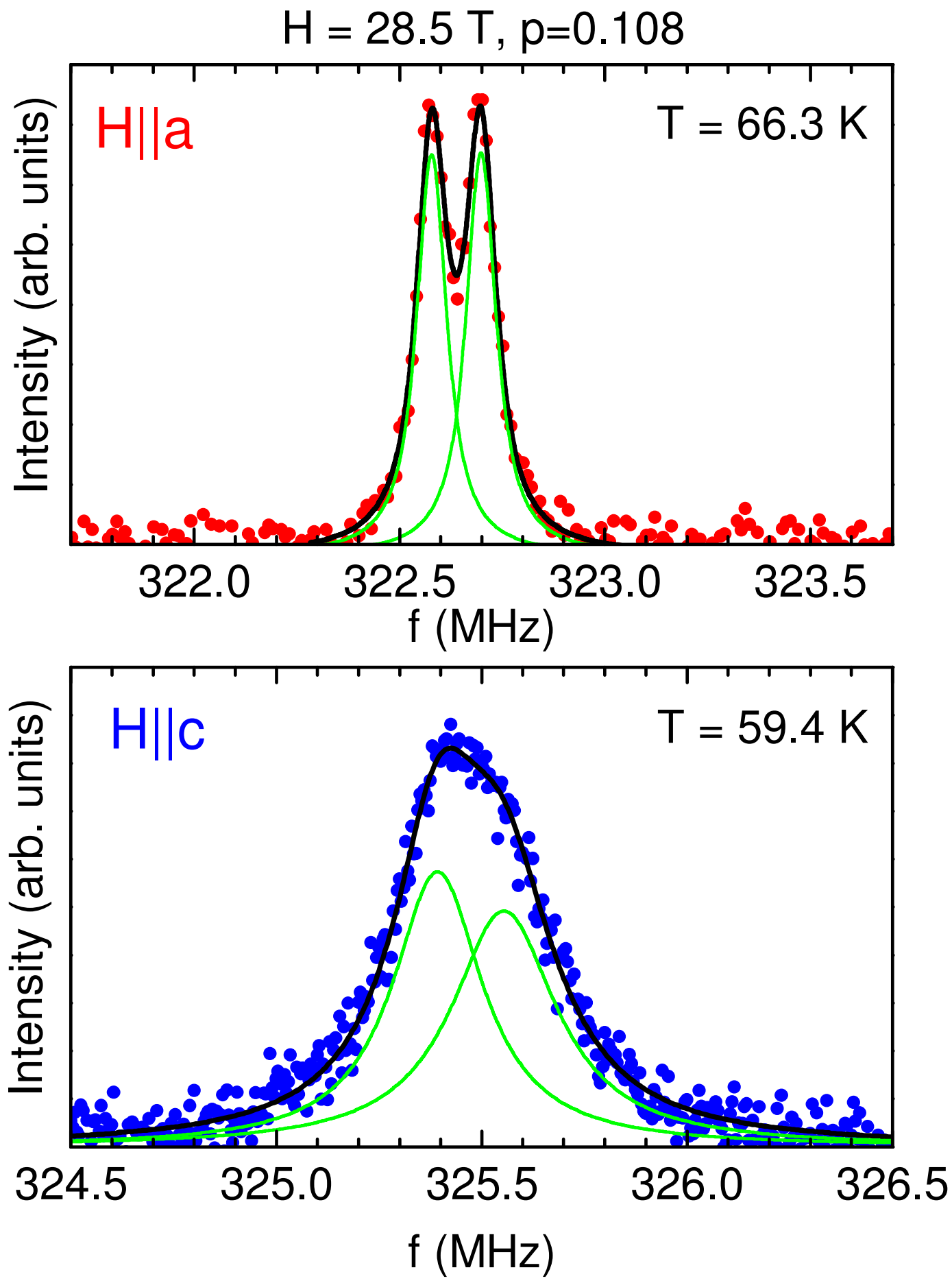

**Figure S2. Anisotropic broadening of the $^{63}$Cu central line of $YBa_2Cu_3O_{6.54}$ (p=0.108).** The green lines represent the Cu2E and Cu2F sites. The larger width of each line for H||*c* (0.27 and 0.31 MHz) than for H||*ab* (0.085 MHz) is evidence that the hyperfine field is staggered and aligned along the field direction.

## III. Hyperfine shifts in the charge-ordered state

In the stripe ordered phase, we assume that the local susceptibility is $\chi_0$ at 2E sites, $\chi_0+\delta\chi$ at 2Fa sites, and $\chi_0-\delta\chi$ at 2Fb sites (implicitly, the local susceptibility is uniform along stripes). Then, the spin contribution to the magnetic hyperfine shift of each site is:

$$Ks(2E)=Acc\chi_0+2B\chi_0+B(\chi_0+\delta\chi)+B(\chi_0-\delta\chi)=(Acc+4B)\chi_0 \approx 0$$

$$Ks(2Fa)=Acc(\chi_0+\delta\chi)+2B(\chi_0+\delta\chi)+2B\chi_0=(Acc+4B)\chi_0+(Acc+2B)\delta\chi \approx -2B\ \delta\chi$$

$Ks(2Fb)=Acc(\chi_0-\delta\chi)+2B(\chi_0-\delta\chi)+2B\chi_0=(Acc+4B)\chi_0-(Acc+2B)\delta\chi \approx 2B\ \delta\chi$

The orbital shift does not show any appreciable doping dependence in $YBa_2Cu_3O_y$, thus it does not contribute to the total shift difference, which reads $\Delta K_{cc}=|K(2Fa)-K(2Fb)|= 4B\ \delta\chi_{cc}$. From our NMR spectra we determine:

$$\Delta K_{cc} = 4B\ \delta\chi_{cc}=0.11\ \% \quad (3)$$

For a doping level of p≈0.16 hole/Cu in $YBa_2Cu_3O_y$, $K_{ab}$(300 K)=0.37 % (39), and for p=0.12 $K_{ab}$(300 K)=0.23 % (40). This gives $\Delta K_{ab}/\Delta p = (A_{ab}+4B)\ \Delta\chi_{ab}/\Delta p = 3.5\ \%$ per hole. Using $A_{cc}$, $A_{ab}$, B and $\chi_c/\chi_{ab}\approx 1.168$ values from (36), and $A_{cc}\approx -4B$, we obtain

$$4B\ \Delta\chi_{cc}/\Delta p = 4.0\ \%\ \text{per hole} \quad (4)$$

A recent $^{63}$Cu NMR work of G.-q. Zheng and coworkers (41) has shown that in magnetic fields strong enough to suppress superconductivity in $Bi_2Sr_{2-x}La_xCuO_{6+\delta}$, the relative doping dependence of $K_{cc}$ in the low temperature limit is the same as that at T=300 K, namely for two different doping levels 0.09<p1,p2<0.19, K(300 K,p1)-K(300 K,p2)=K(0 K,p1)-K(0 K,p2). Thus, we assume that the relation $4B\ \Delta\chi_{cc}/\Delta p = 4.0$ established from room temperature data in $YBa_2Cu_3O_y$ holds at low temperatures as well, for our high field data.

Combining (3) and (4), we obtain that the experimental hyperfine shift difference of 0.11 % between Cu2Fa and Cu2Fb implies a charge density difference of 0.028 hole, in remarkable agreement with $\delta p=0.03$ hole inferred from the quadrupole splitting of the Cu2F line.

Even if the presented analysis is somewhat crude, the consistency of the results shows that the splitting seen on the central line can be explained by a difference in the spin (not orbital) contributions to the hyperfine shift. This difference arises from the spatial modulation of the spin polarisation in the ordered striped state.

## IV. Signal overlaps

At 9 and 15 Tesla, the planar Cu2 central line overlaps with the Cu1F signal from oxygen-filled chains (Fig. S3). Because the Cu1F signal is extremely broad, its amplitude at the Cu2 resonance frequency is small (5 % maximum for both samples) and it does not affect our Cu2 relaxation rate or line shape measurements. Below 20 K, due to extreme broadening and short $T_2$, the Cu1F signal becomes invisible. This Cu2-Cu1F overlap is absent at all temperatures in high fields (28-33 T). Above ~30 T, on the other hand, the $^{63}$Cu2 central line may overlap with the low frequency $^{65}$Cu satellite (and the $^{65}$Cu2 central line with the high frequency $^{63}$Cu satellite), depending on the field values. All the high field results presented in this work are taken in conditions for which this overlap does not occur, except for the relaxation rate measurements at 33.5 T in $YBa_2Cu_3O_{6.67}$ (p=0.108). However, these results are again unaffected by this overlap, because the satellite signal is small and the rate values are completely consistent with those at close field values for which no overlap occurs.

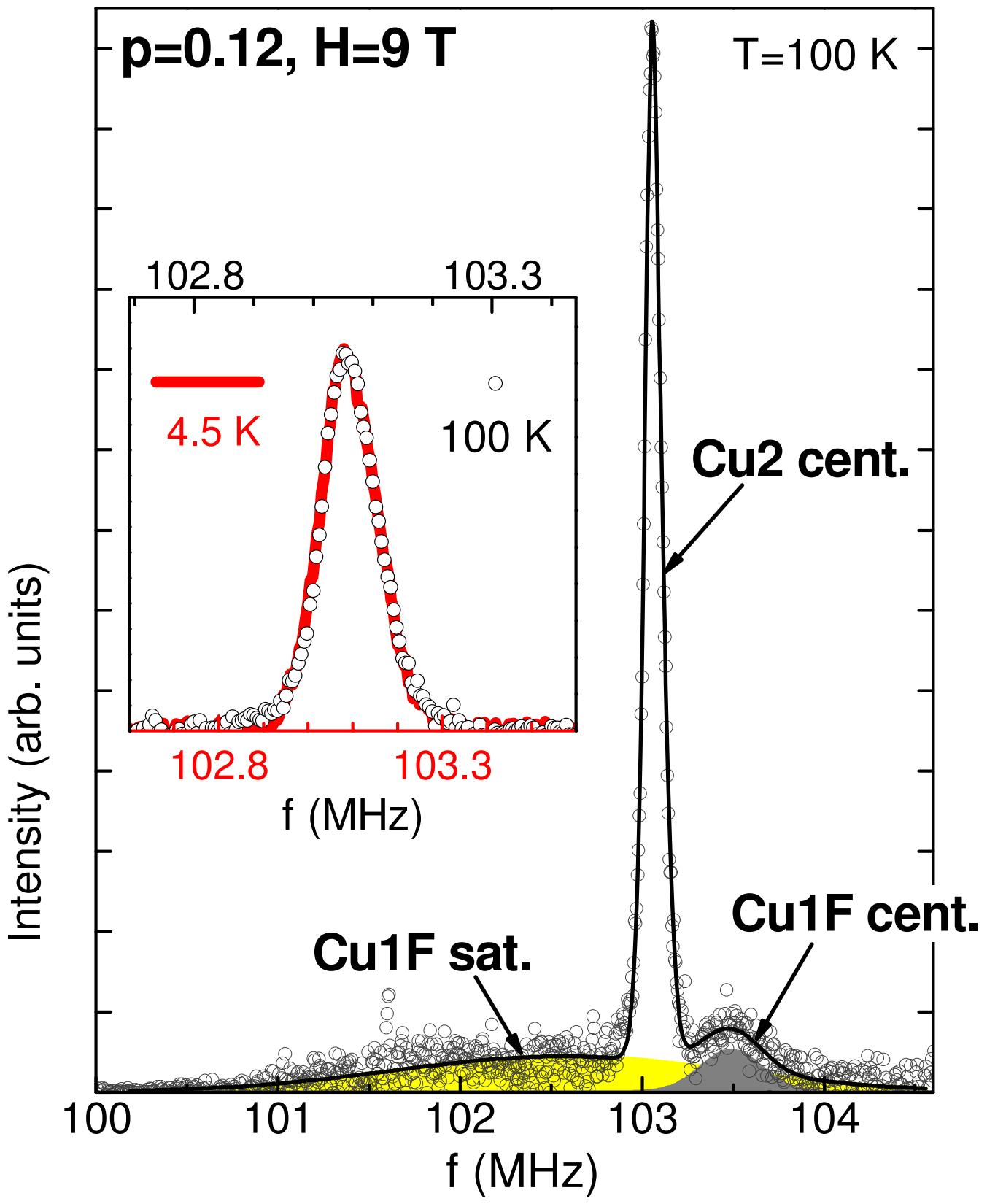

**Figure S3. Low field Cu2 central line of $YBa_2Cu_3O_{6.67}$.** The two Cu1F satellites (broad yellow area) are unresolved and are both on the same side of the central line because the asymmetry parameter of the electric field gradient is close to 1 for this site. Inset: Comparison of the Cu2 central lines at two different temperatures (a background has been subtracted from the data at 100 K, but not at 4.5 K). Aside from a slight position difference, the line shape is remarkably insensitive to temperature.

## V. Low field NMR spectra

At high temperatures and/or low fields (9-15 T), our spectra are consistent with previous studies on similarly-doped $YBa_2Cu_3O_y$. In particular, for y=6.5, the Cu2E and Cu2F sites can directly be resolved at 9 T (Fig. S3), in agreement with Ref. 10.

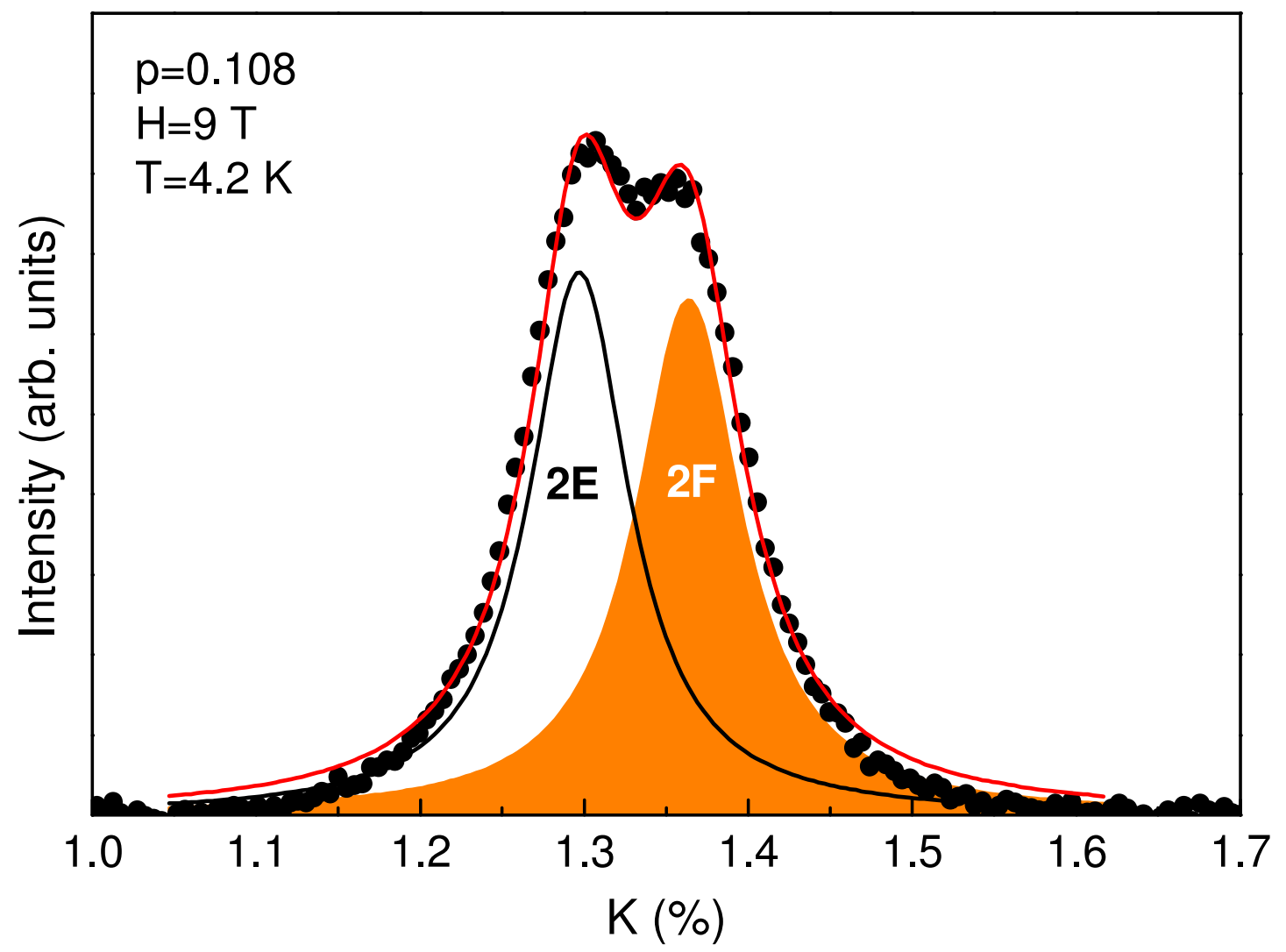

**Figure S4 | $^{63}$Cu NMR central line of $YBa_2Cu_3O_{6.54}$ (ortho-II, p=0.108).** The Cu2E (planar sites below empty chains) and Cu2F (planar site below oxygen-filled chains) sites are directly resolved.

## VI. High-field NMR spectra

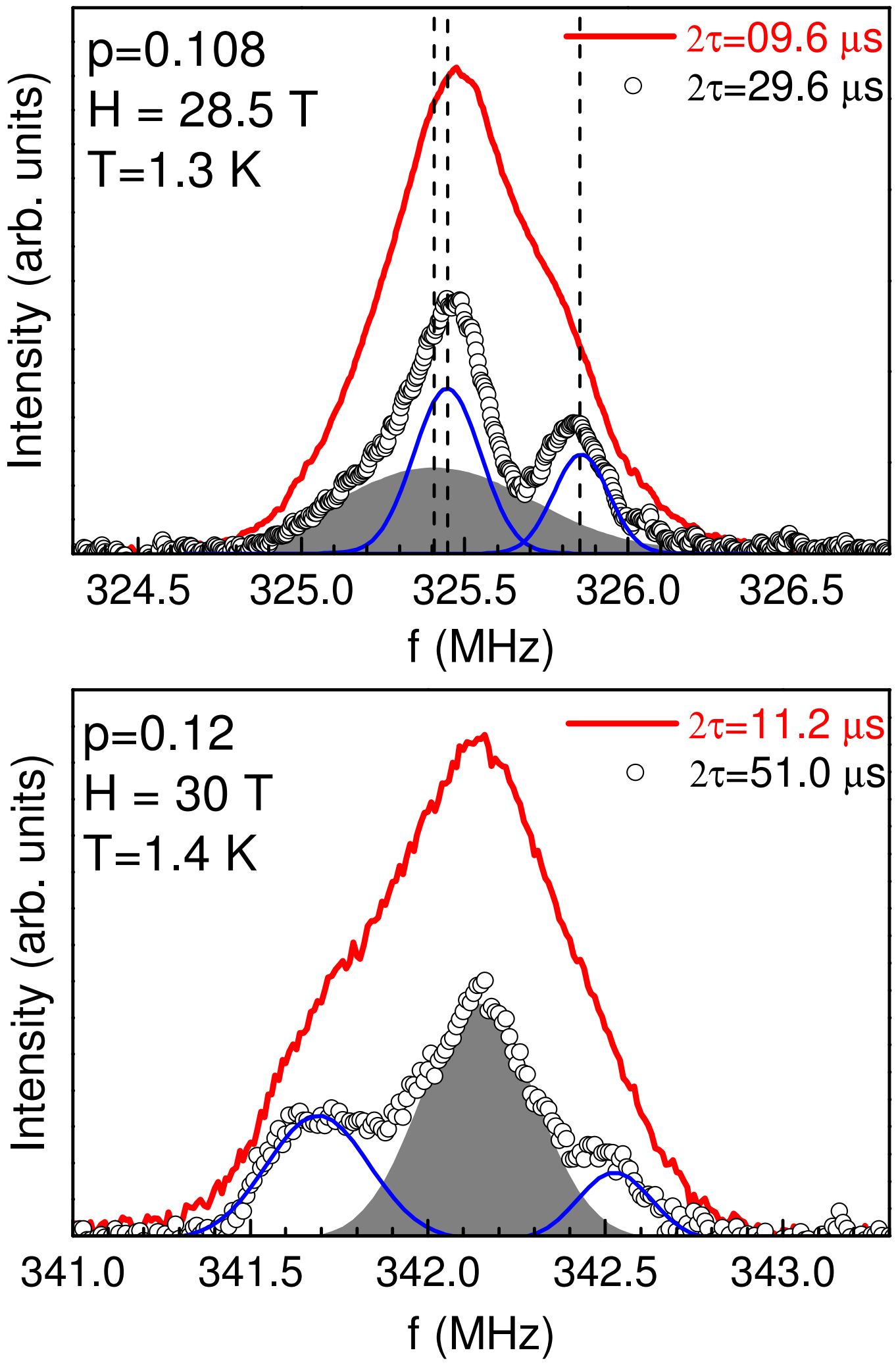

**Figure S5 | Revealing the underlying central line positions with "long $\tau$ spectra".** Although the standard NMR spectra (in red) show only shoulders at low temperature and high fields, using longer values of the total time $\tau$ between $\pi/2$ and $\pi$ pulses of the spin-echo sequence narrows the lines and reveals the hitherto unresolved lines (Cu2E in grey, Cu2Fa,b in blue). The effect is due to the fact that long-$\tau$ spectra eliminate the signal from nuclei with shorter $T_2$ values, which mostly contribute to the edges of each of the lines (see sections II and VI). Because of the inhomogeneity of $T_2$ and $\alpha$ across each line (See Fig. S11) and possible differences between the three lines, the relative intensities in this procedure may be different from real intensities. Actually, the short and long $\tau$ are not on the same scale: The signal amplitude for long $\tau$ spectra represents only about 1/10$^{th}$ of the signal at short $\tau$. Thus, no conclusions regarding the relative intensities or widths of the different lines should be drawn from such spectra. For p=0.108, the central line positions marked by dashed lines are deduced from the positions of the quadrupole satellites.

| | $K_c$ (60 K) | $K_c$ (1.3 K) | $\nu_Q$ (60 K) | $\nu_Q$ (1.3 K) |
|---|---|---|---|---|
| Cu2E | 1.31(2) % | 1.32(1) % | 27.37(6) MHz | 27.40(4) MHz |
| Cu2Fa | 1.36(2) % | 1.32(1) % | 30.47(6) MHz | 30.37(4) MHz |
| Cu2Fb | | 1.43(1) % | | 30.70(4) MHz |

**Table S1 | NMR parameters for $YBa_2Cu_3O_{6.54}$ (p=0.108).** The magnetic hyperfine shift (K) and quadrupole frequency ($\nu_Q$) values are obtained by matching the experimental positions with those calculated by exact diagonalisation of the nuclear-spin Hamiltonian.

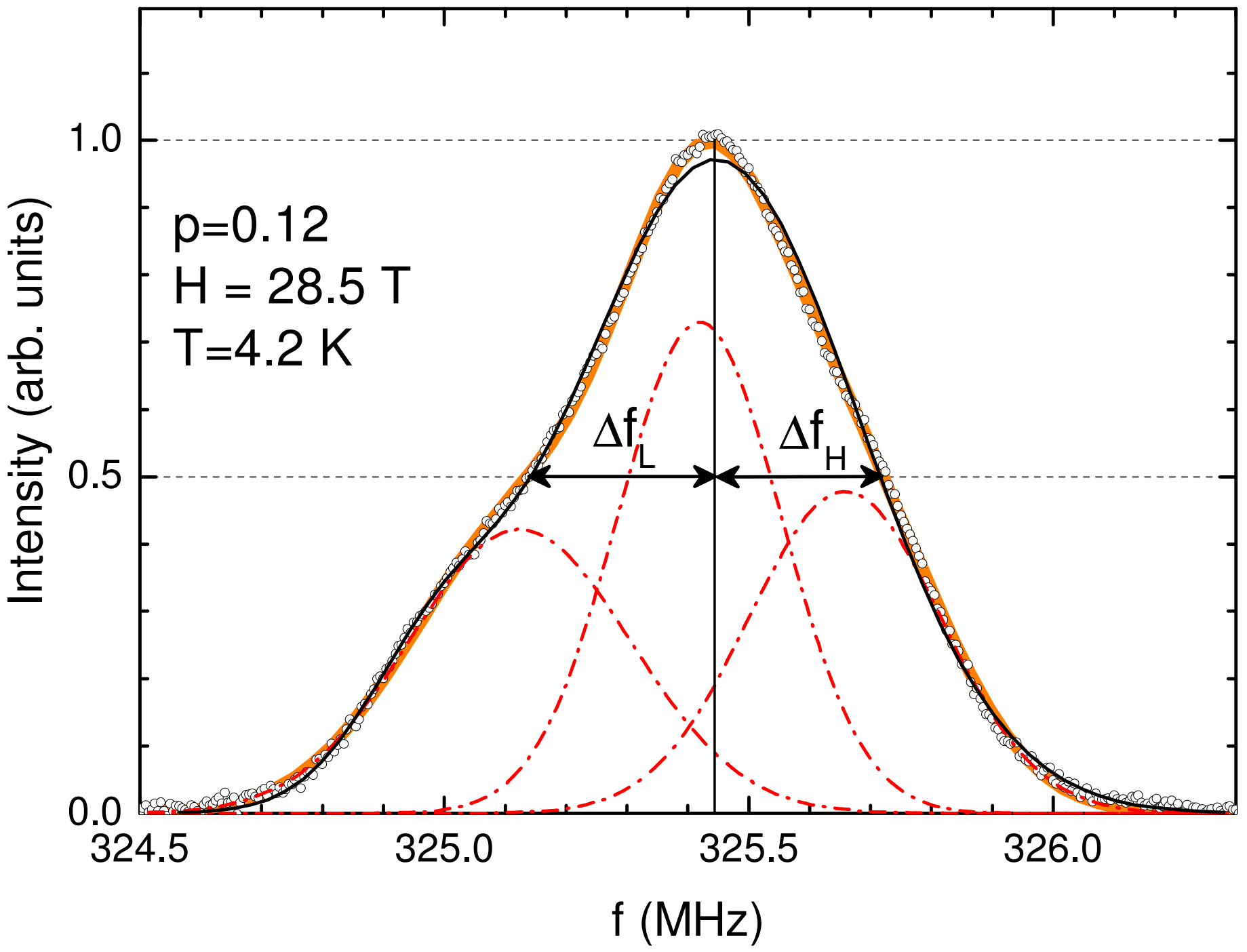

**Figure S6 | Example of high field $^{63}$Cu central lines for $YBa_2Cu_3O_{6.67}$ (p=0.12).** This spectrum (open circles) shows the typical signal-to-noise ratio achieved at low temperatures and high fields, and defines the half-widths $\Delta f_L$ and $\Delta f_H$ used to characterise the asymmetry A of the central lines in Figs. S8 and S10: $A=(\Delta f_L-\Delta f_H)/(\Delta f_H+\Delta f_L)$. The fit to three Gaussian lineshapes (orange) is visibly better than the fit to two Gaussians (black). Lorentzian lineshapes do not fit. The existence of three distinct resonance positions for this sample is independently confirmed by spectra in Fig. S5.

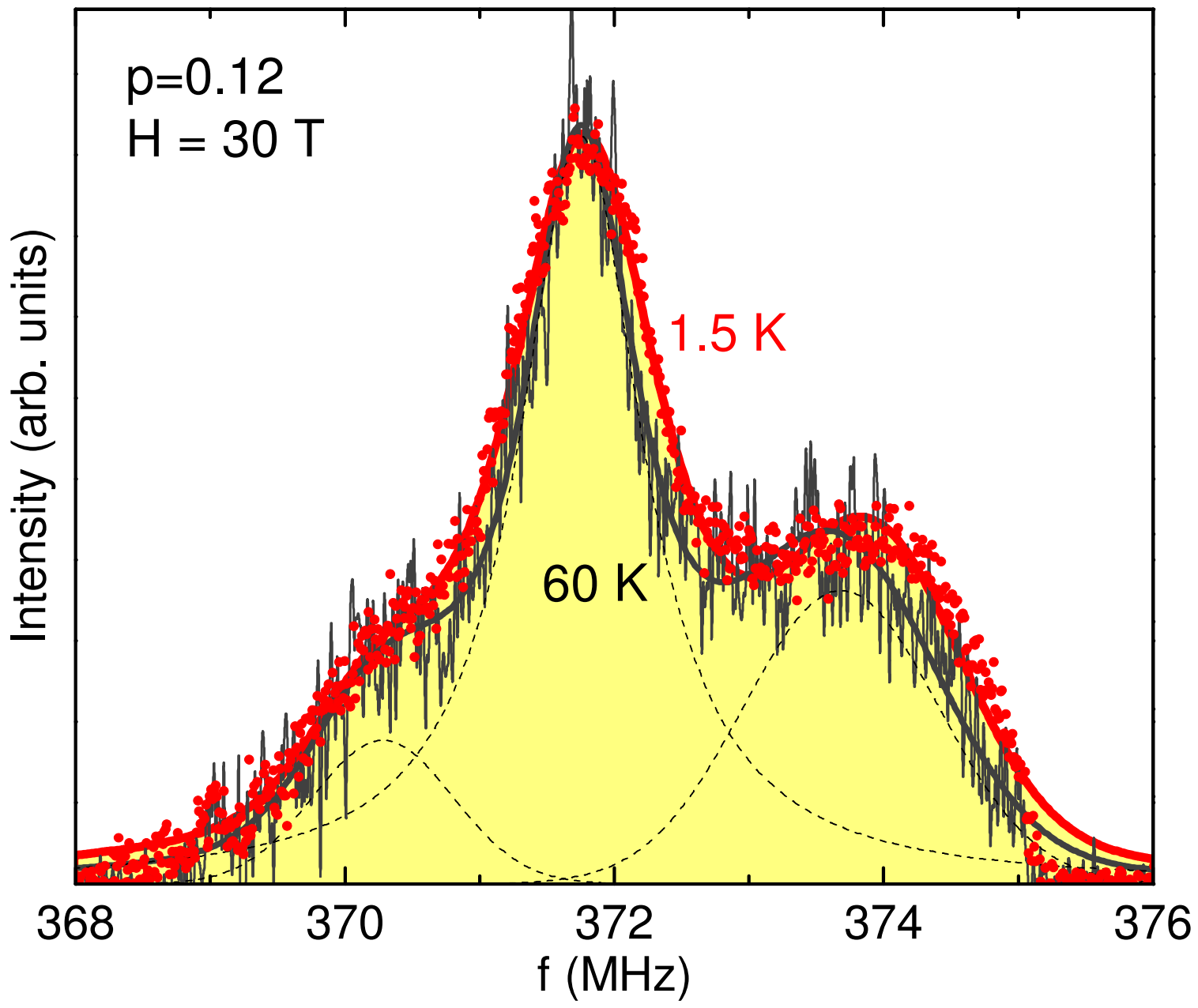

**Figure S7 | High frequency quadrupole satellites for $YBa_2Cu_3O_{6.67}$ (ortho VIII, p=0.12).** The ortho VIII structure is more complex and not as well ordered as ortho II. Unlike the ortho II sample (p=0.108) with only two separated sites, there are here at least three different values of the quadrupole frequency at planar $^{63}$Cu sites (See table S2). Therefore, the modifications of $\nu_Q$ on cooling at high field are difficult to identify. Although the spectrum is intrinsically broad, some extra broadening at 1.5 K can still be observed.

| | $K_c$ *(60K)* | $\nu_Q$ *(60K)* |
|---|---|---|
| Cu2a | 1.3(1) % | 27.8(2) MHz |
| Cu2b | 1.3(1) % | 29.3(2) MHz |
| Cu2c | 1.4(1) % | 31.2(2) MHz |

**Table S2 | NMR parameters for $YBa_2Cu_3O_{6.67}$ (p=0.12).** The magnetic hyperfine shift (K) and quadrupole frequency ($\nu_Q$) values are obtained by matching the experimental positions with those calculated by exact diagonalisation of the nuclear-spin Hamiltonian.

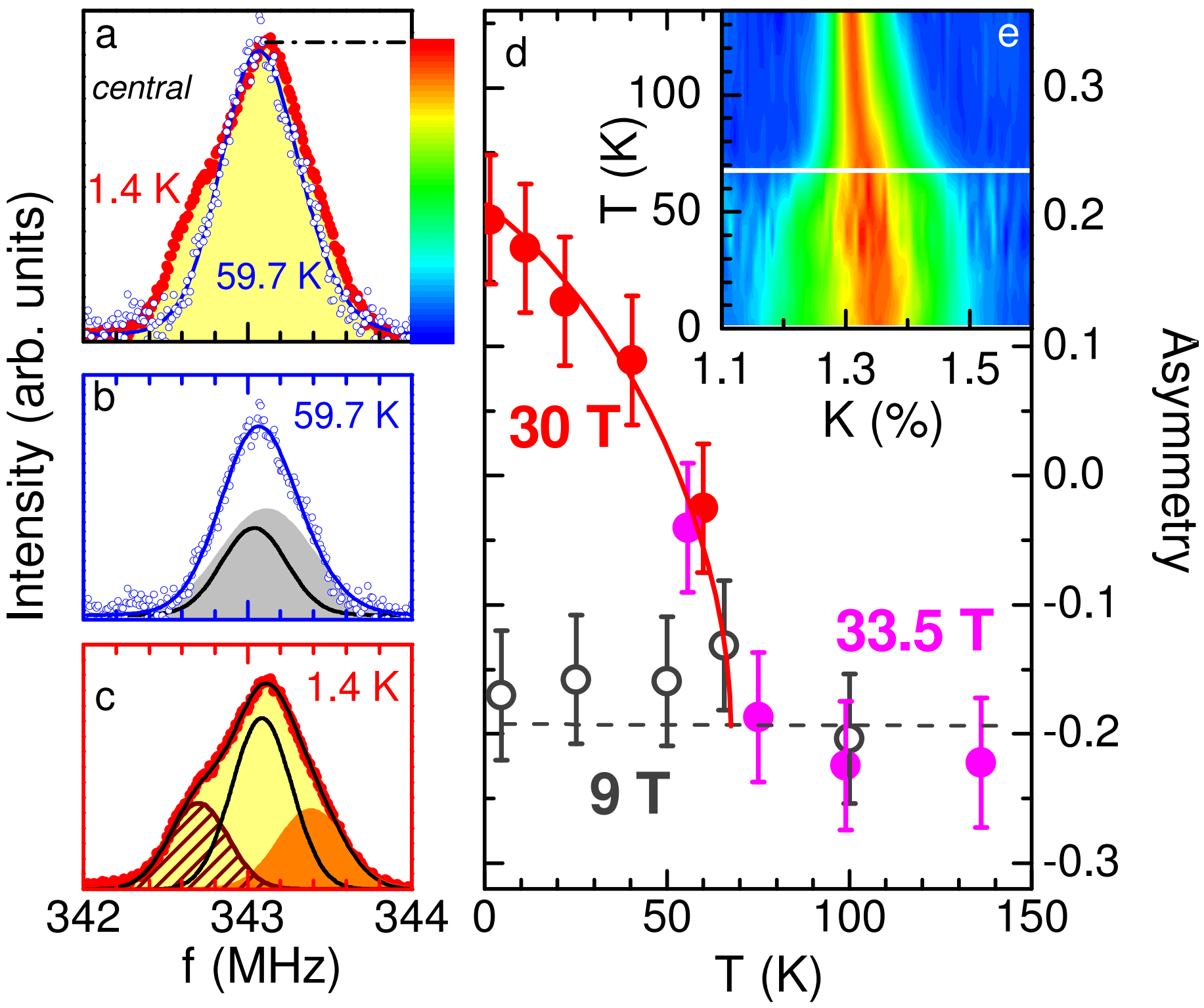

**Figure S8 | High field NMR spectra of $YBa_2Cu_3O_{6.67}$ (ortho-VIII, p=0.12). a**, comparison of the $^{63}$Cu central lines (-1/2,1/2 transitions) at $T \approx T_{charge}$ (blue) and $T << T_{charge}$ (red) in a magnetic field of 30 T. **b**, decomposition of the high T spectrum into two sites. Although there are three different sites from the electric-field-gradient point of view (see Fig. S7), the ortho-VIII structure contains only filled and empty chains which should give rise to only two different magnetic hyperfine fields, as in the ortho-II compound, given the local character of the hyperfine interaction (on-site and first neighbours polarisation). From the magnetic hyperfine shift point of view, Cu2E (black line) and Cu2F (grey) sites might thus be defined. **c**, Decomposition of the low T spectrum (red) into Cu2E (black line), Cu2Fa (hatched) and Cu2Fb (orange) sites. The two latter lines originate from the splitting of high temperature Cu2F line. **d**, temperature dependence of the asymmetry of the central line $A=(\Delta f_H-\Delta f_L)/(\Delta f_H+\Delta f_L)$ where $\Delta f_H$ and $\Delta f_L$ are defined as in Fig. S6. Note that the asymmetry is opposite to that in p=0.108 (Fig. S10). This could be due to slightly different position of the average Cu2F position with respect to Cu2E in the charge-ordered state (For instance a rotation of the principal axis of the electric field gradient tensor coupled to a non-zero quadrupole asymmetry parameter would produce a second-order quadrupolar shift). These fine details of the qualitative interpretation cannot easily be solved experimentally because changing the field orientation changes (and eventually cancels) the Cu2Fsplitting. The asymmetry is a direct consequence of the shoulder on the low frequency side, itself due to the splitting of the Cu2F line and is clearly field-induced. Lines are guides to the eye. **e**, NMR intensity vs. T and hyperfine shift K from the data at 30 and 33.5 T. The horizontal white line marks the temperature (~67 K) at which asymmetry shows a marked change.

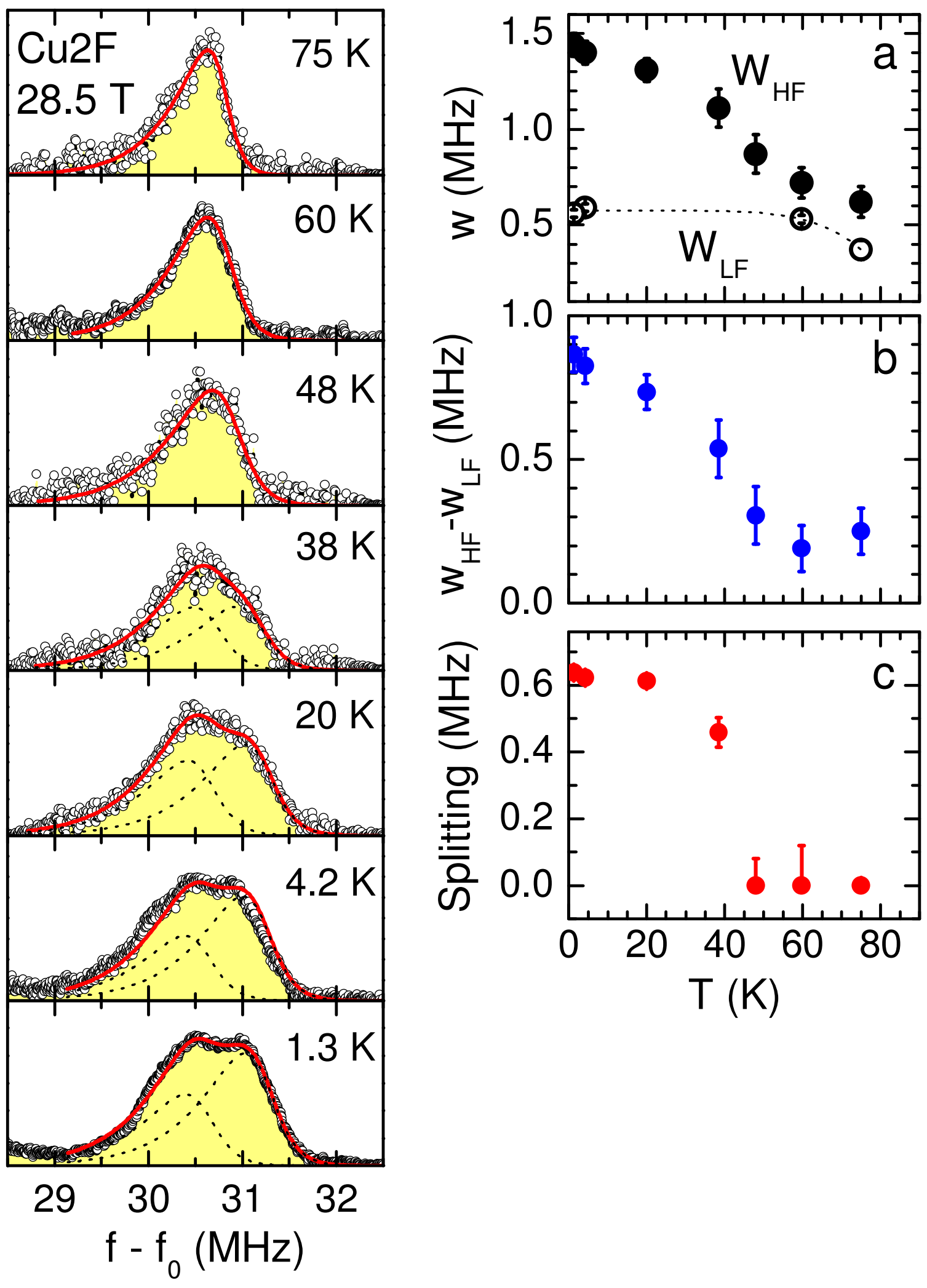

**Figure S9 | $^{63}$Cu(2F) quadrupole satellite splitting for $YBa_2Cu_3O_{6.54}$ (ortho II, p=0.108).**
**Left**: High frequency quadrupole satellites as a function of temperature. $f_0$ is equal to 326.0 MHz. The continuous line is a fit using two lines of equal widths but different amplitudes, in order to account for the shorter $T_2$ of the lowest frequency line (which has been independently checked to be consistent with the intensity ratios of the fit here). Each line has the same intrinsic (quadrupolar) asymmetry as the line at 75 K.
**Right**: **a**, width (w) of the high frequency (HF) and low frequency (LF) satellites. Above $T_{charge}$=50 K, both widths depend on temperature because of the magnetic broadening (also observed on the central line). Below $T_{charge}$, $w_{LF}$ saturates because the magnetic broadening saturates (consistent with analysis of the central line at low and high fields) and the quadrupolar splitting cancels on this satellite. $w_{HF}$, on the other hand, increases sharply because of the quadrupolar splitting. **b**, $w_{HF}$-$w_{LF}$ represent the quadrupolar-broadening contribution to the total broadening of the HF satellite. **c**, Total splitting obtained from the fits shown in the left column (The absence of modification on the low frequency satellite indicates that the quadrupolar and magnetic hyperfine contribute equally to this total splitting). Despite a slight difference in the T dependence, both $w_{HF}$-$w_{LF}$ and the splitting indicate an onset temperature of 50±10 K for the charge order. Error bars at 75, 60 and 48 K in panel c indicate the splitting if a fit with two lines is forced.

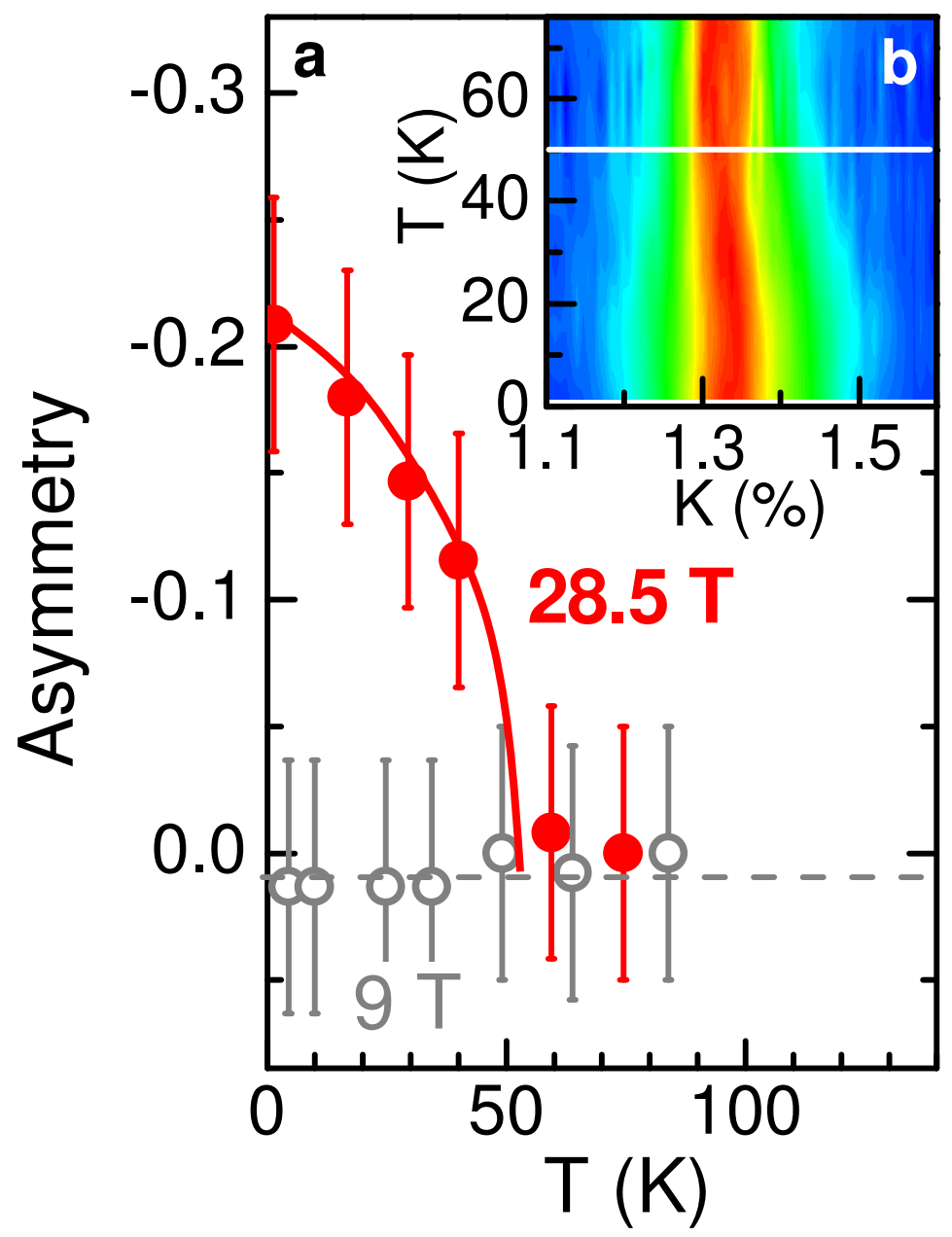

**Figure S10 | Asymmetry of the $^{63}Cu(2)$ central line for $YBa_2Cu_3O_{6.54}$ (ortho II, p=0.108).**
**a**, The asymmetry (See definition in Figs. S6) is a direct consequence of the splitting of the Cu2F central line and is clearly field-induced below 50 K. Lines are guides to the eye. Inset, **b**, NMR intensity vs. T and hyperfine shift K from the data at 28.5 T.

.

## VII. Spin-spin relaxation rate $1/T_2$

As shown in Figure 3, the spin-echo decay is described by a stretched exponential form below ~$T_{charge}$. Part of this behaviour is due to the appearance of a distribution of $T_2$ values across the NMR line, *i.e.* $T_2$ becomes spatially inhomogeneous (Fig. S11).

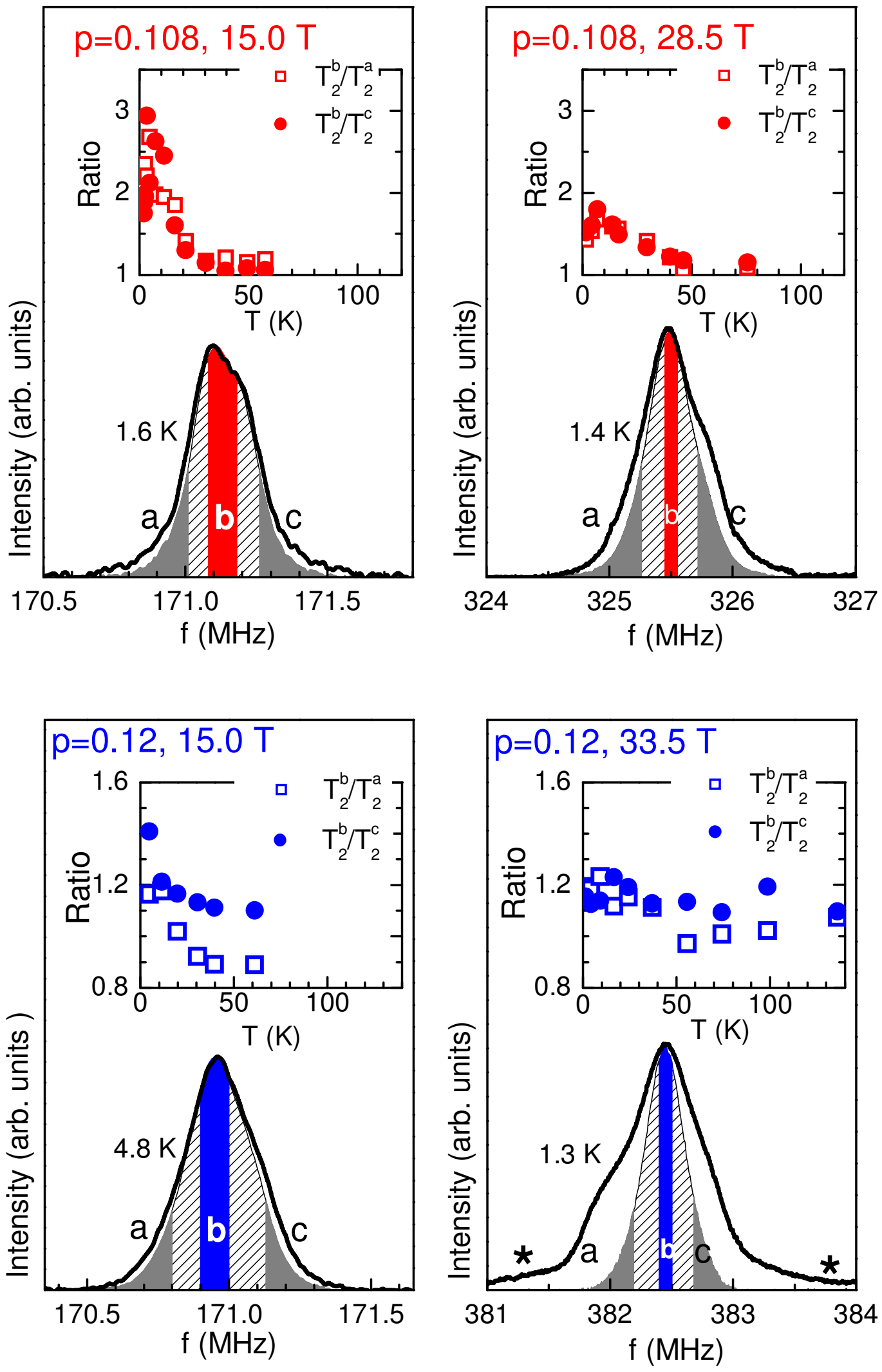

**Figure S11 | $T_2$ inhomogeneity for $YBa_2Cu_3O_{6.54}$ (p=0.108) and $YBa_2Cu_3O_{6.67}$ (p=0.12).** a, b and c areas represent the zones over which the signal is integrated, in order to define $T_2^a$, $T_2^b$ and $T_2^c$ (insets). These grey and blue/red areas, plus the dashed areas, represent the signal obtained from a Fourier transform of the (half) spin-echo at a single frequency in the $T_2$ experiments. The whole spectrum (black thick line) is obtained in a separate experiment, by adding frequency-shifted Fourier transforms. $T_2$ is shorter on the edges of the spectrum (a and c areas), *i.e.* at positions where the amplitude of the staggered spin polarisation is the largest. The inhomogeneity appears to be lower at high fields because a smaller portion of the line is covered by the pulses, and/or because of a possible extra inhomogeneity in the vortex solid phase at low fields. At 33.5 T for p=0.12, the broad background marked with a star arises from the $^{65}Cu$ quadrupole satellites (See supplementary section IV). Note that only $T_2$ ratios are plotted, so the fact that $1/T_2$ is higher at higher fields does not appear here.

However, the value of the stretching exponent α at high field does not vary when the frequency window of the analysis is narrowed (Figure S12). This shows that the stretched form is intrinsic and primarily arises from the appearance of an exponential decay in addition to the usual Gaussian decay which dominates at temperatures above $T_{charge}$. The presence of an exponential decay is further confirmed by the isotope ratio of $T_2$ values (see below).

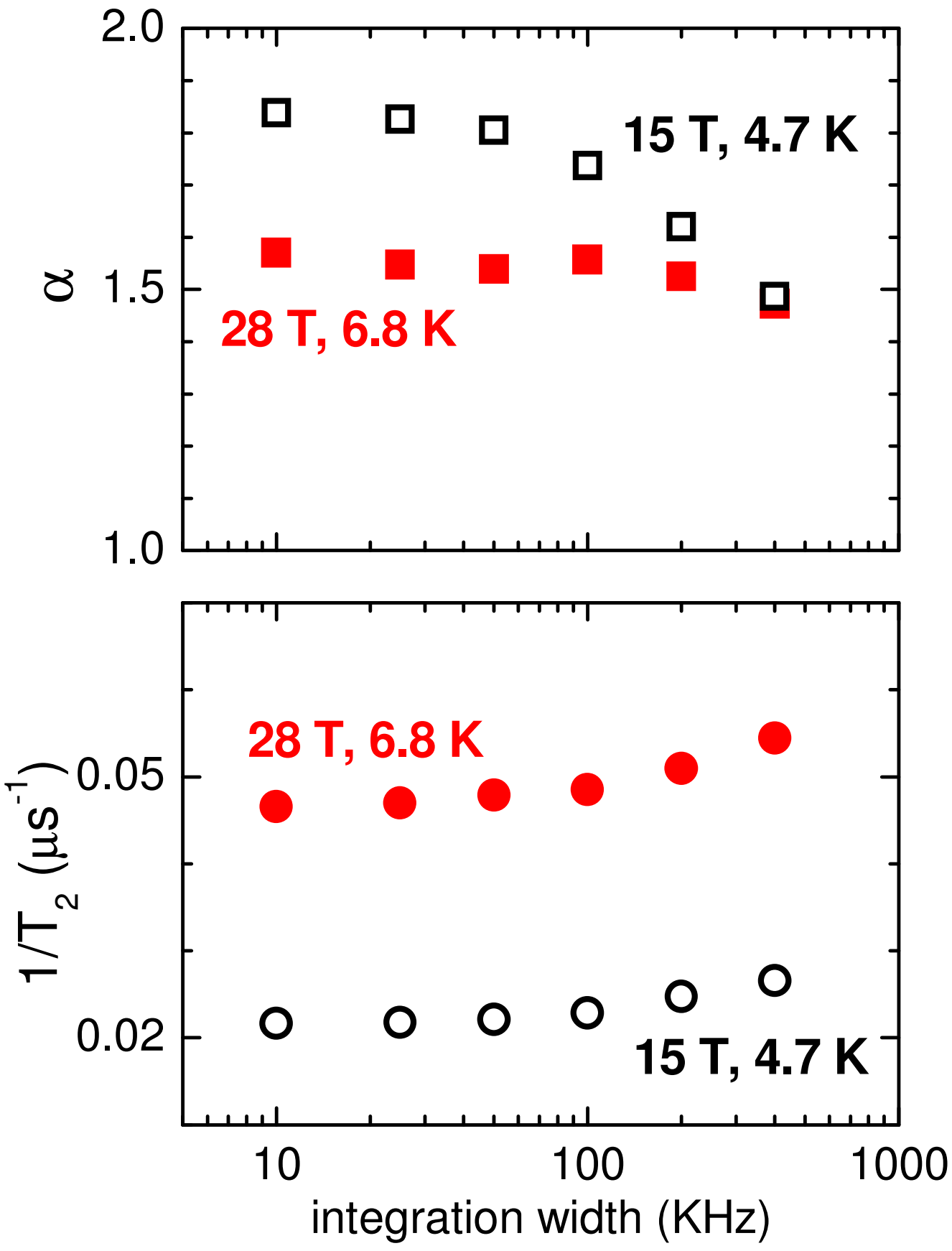

**Figure S12 | Combination of exponential and Gaussian contributions to the spin-echo decay for $YBa_2Cu_3O_{6.54}$ (p=0.108).** The stretching exponent α (see Fig. 3) does not change when reducing the frequency width over which the signal is integrated, especially at high fields.

Because of the inhomogeneous character of the spin polarisation (staggered spin polarisation at all temperatures combined with further low temperature modulations due to stripe-order at high fields and due to the vortex solid state at low fields, as well as possible spatial inhomogeneity of unknown typical length scale), there is no reliable way to disentangle the exponential and Gaussian contributions to the echo decay. A fit including both $T_{2E}$ and $T_{2G}$ decays together with a stretching exponent α (accounting for the inhomogeneity) as free parameters does not yield consistent results. However, different analysis procedures (product or sum of exponential and Gaussian decays) with either $T_{2G}$ or α being fixed, all lead to the conclusion that an exponential decay appears below ~$T_{charge}$, with $T_2$ values of ~100 μs close to $T_{charge}$, shortening to ~10 to 25 μs at low T (See Fig. S13 for one such analysis).

The exponential decay is typical of a relaxation process governed by the fluctuations of the longitudinal component of the local field $h_z$ (Ref. 42) in the limit $\gamma <h_z^2>^{1/2}\tau_c <<1$ where $\tau_c$ is the correlation time of the fluctuations defined as $<h_z(t)h_z(0)> \propto \exp(-t/\tau_c)$ and

$$1/T_{2E} = \gamma^2 <h_z^2> \tau_c \qquad (3)$$

Here, the magnetic nature of the fluctuations is indeed ascertained by the values of the isotope ratio $^{65}T_2^{-1}/^{63}T_2^{-1}$=1.09 for p=0.12 (T=11 K) and 1.05 for p=0.108 (T=1.4 K). These values are consistent with the ratio $(^{65}\gamma/^{63}\gamma)^2$=1.148 expected from equation (3), rather than with the ratio of quadrupole moments expected for electric-field-gradient fluctuations $(^{65}Q/^{63}Q)^2$=0.859 or with the ratio $(0.31/0.69)^{0.5}(^{65}\gamma/^{63}\gamma)^2$=0.77 expected in the case of a purely Gaussian decay due to magnetic relaxation.

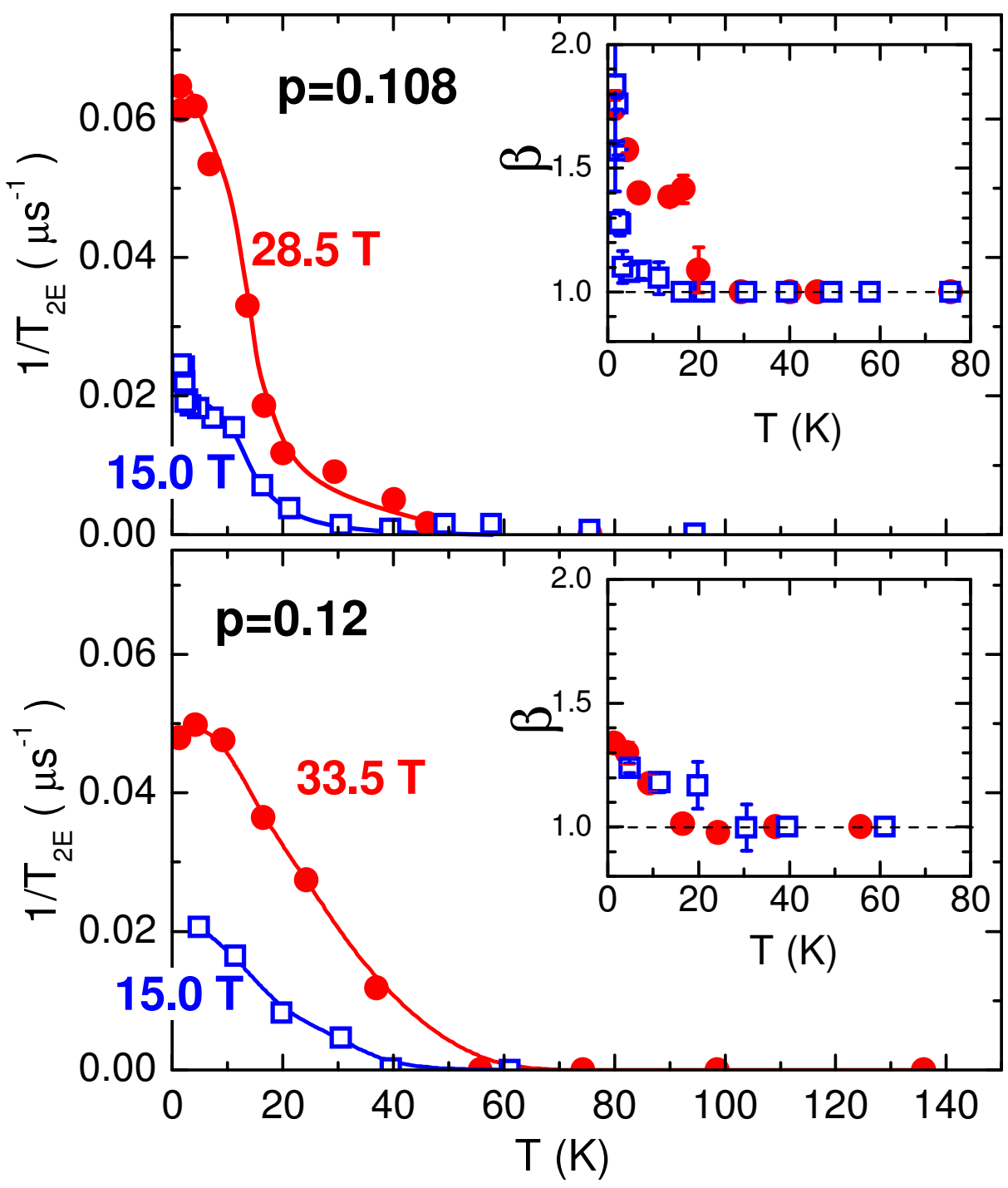

**Figure S13 | Exponential contribution to the spin-echo decay.** Here, the raw data were fit to the formula $s(t)=m*\exp(-1/2*(t/T_{2G})^2)*\exp(-t/T_{2R})*\exp(-(t/T_{2E})^{\beta})$, with $T_{2E}$ and β as the only free parameters below $T_{charge}$: For $T<T_{charge}$, $T_{2G}$ was fixed to the value at T slightly higher than $T_{charge}$ for each field and sample (Fig. 3). The Redfield contribution is given by $T_{2R}=(3+R)T_1^{-1}$ where R=3.6 is the $T_1$ anisotropy ratio (Refs. 43,44). Lines are guides to the eye. The increase of β at low temperatures is due to spatial inhomogeneity and/or to the condition $\gamma <h_z^2>^{1/2}\tau_c <<1$ being no longer met.

These fluctuations cannot be related to vortex motion since their onset temperature is considerably higher than the vortex-lattice melting transition (even above $T_c$ for p=0.12 at 33.5 T). The shorter $T_2$ values on the edges of each line where the spin polarisation is larger (See Fig. S11) shows that the fluctuations actually arise from the staggered $Cu^{2+}$ spins. In these conditions, the typical values of $T_{2E}$ imply correlation times in the range $10^{-9}$–$10^{-12}$ s for any value of $g^2\langle S_z^2\rangle$ lower than ~0.1 $\mu_B^2$ (Figure S14). Thus, for any value of the spin polarisation compatible with the Cu NMR linewidth, the staggered moments appear to be frozen on the timescale of a cyclotron orbit $1/\omega_c \approx 10^{-12}$ s.

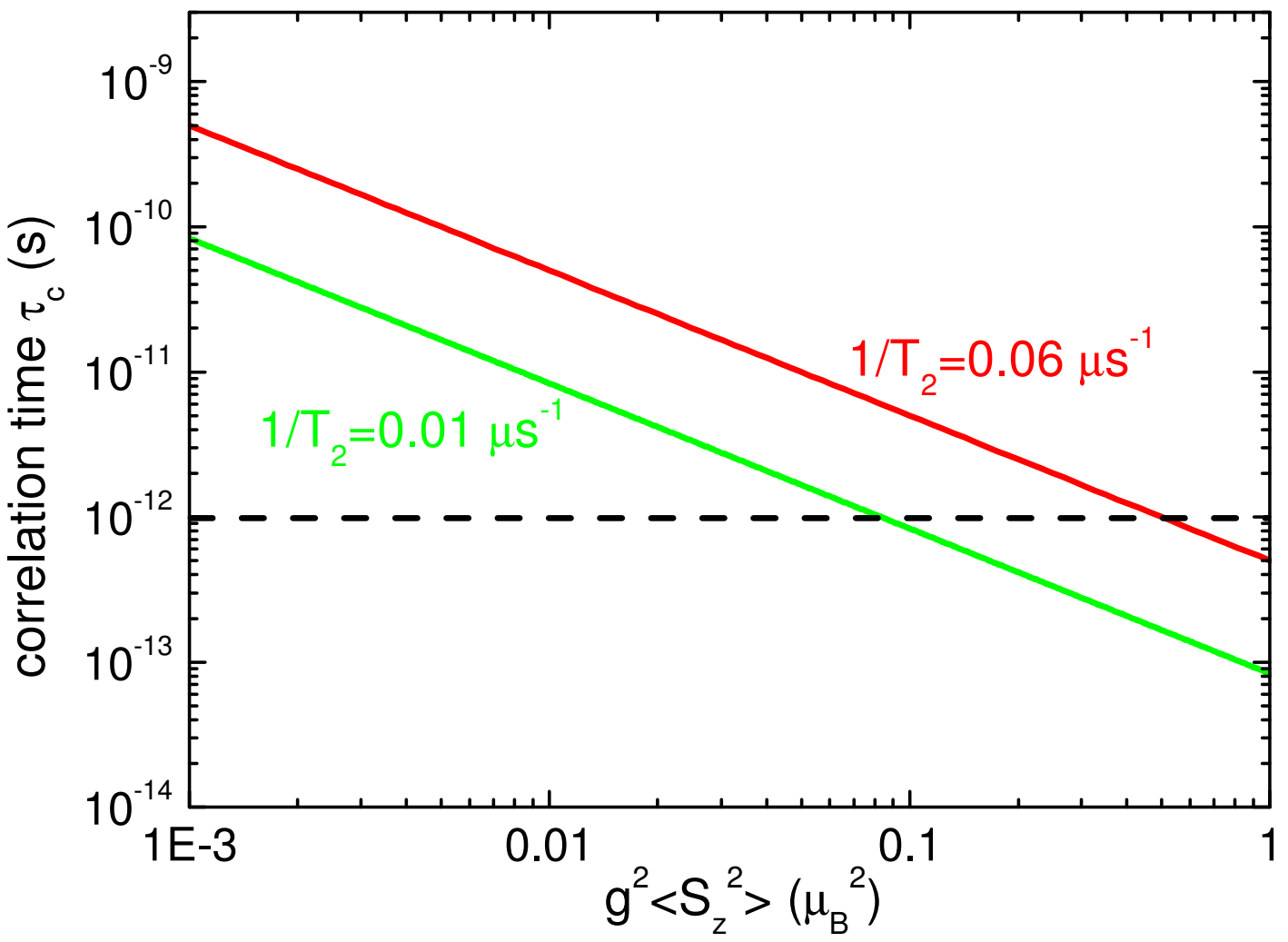

**Figure S14 | Correlation times deduced from $T_2$ measurements.** $\tau_c$ values are calculated from Eqn. (3) for two typical values of $1/T_{2E}$ (see Fig. S13 and text). The dashed line represents the timescale of a cyclotron orbit $1/\omega_c \approx 10^{-12}$ s.

## VIII. Spin-lattice relaxation rate $1/T_1$

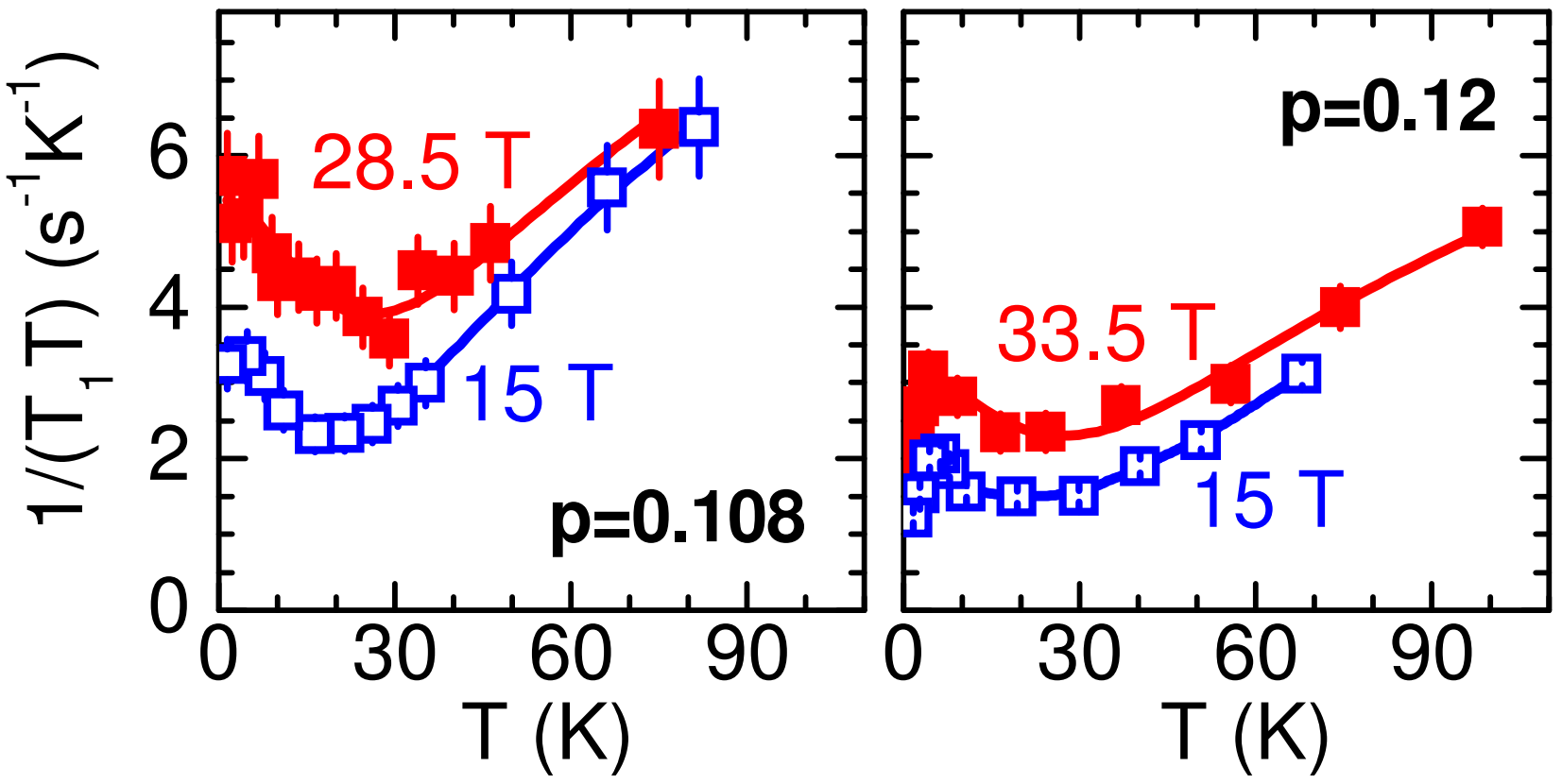

**Figure S15 | Spin-lattice relaxation rate divided by T.** In contrast with the monotonous decrease of $1/T_1$ on cooling, the quantity $1/(T_1T)$ shows an upturn at low temperature, eventually followed by a decrease for the p=0.12 sample (For p=0.108, the saturating value of $1/T_1T$ around 1.5 K suggests that a similar decrease may occur at lower T). Since the temperature at which $1/T_1T$ is maximum shifts to lower temperature on increasing the field, the decrease should be associated with superconductivity (vortex physics or pairing). Analysis of these low temperature/field phenomena are beyond the scope of this work (In any event, the peak occurs at much lower temperature than the stripe ordering/Fermi surface reconstruction onset). Note that the overall increase of $1/(T_1T)$ on increasing the field in the superconducting state is due to excitations of nodal quasiparticules, as shown by previous high field NMR measurements in the mixed state of $YBa_2Cu_3O_y$ (45,46).

### Supplementary references